\documentclass[moor,nonblindrev]{informs1}

\usepackage{physics}
\usepackage{mathtools}
\usepackage{tensor}
\usepackage{comment}

\usepackage[mathscr]{euscript}   % Euler Script font:
\newcommand{\half}{\tfrac{1}{2}}

\newcommand{\Rez}{\mathop{\rm Re} \nolimits}

\newcommand{\BoolAnd}{\mathrel{\rm \&}}

\newcommand{\rr}{\sigma}

\newmuskip\pFqmuskip
\newcommand*\pFq[6][8]{%
  \begingroup % only local assignments
  \pFqmuskip=#1mu\relax
  \mathcode`\,=\string"8000
  \begingroup\lccode`\~=`\
  \lowercase{\endgroup\let~}\pFqcomma
  {}_{#2}F_{#3}{\left[\genfrac..{0pt}{}{#4}{#5};#6\right]}%
  \endgroup
}
\newcommand{\pFqcomma}{\mskip\pFqmuskip}

\def\wscl{0.8}
\def\hscl{0.56}

\def\wsclx{1.0}
\def\hsclx{0.7}

\usepackage{natbib}
\NatBibNumeric
\def\bibfont{\small}%
\bibpunct[, ]{[}{]}{,}{n}{}{,}%

\definecolor{myblue}{RGB}{0, 20, 114}
\usepackage[colorlinks=true,breaklinks=true,bookmarks=true,urlcolor=myblue,
     citecolor=myblue,linkcolor=myblue,bookmarksopen=false,draft=false]{hyperref}

\def\EMAIL#1{\href{mailto:#1}{#1}}% When hyperref is used, otherwise outcomment
\usepackage{algorithmic}
\usepackage{algorithm}

\makeatletter
\newcommand{\nextverbatimspread}[1]{%
  \def\verbatim@font{%
    \linespread{#1}\normalfont\ttfamily% Updated definition
    \gdef\verbatim@font{\normalfont\ttfamily}}% Revert to old definition
}
\makeatother

\TheoremsNumberedThrough     % Preferred (Theorem 1, Lemma 1, Theorem 2)
\EquationsNumberedThrough    % Default: (1), (2), ...
\MANUSCRIPTNO{0} % When the article is logged in and DOI assigned to it,
\def\theARTICLETOP{}
\def\theLRHSecondLine{}
\def\theRRHSecondLine{}

\begin{document}
%%%%%%%%%%%%%%%%

% Outcomment only when entries are known. Otherwise leave as is and
%   default values will be used.
%\setcounter{page}{1}
%\VOLUME{00}%
%\NO{0}%
%\MONTH{Xxxxx}% (month or a similar seasonal id)
%\YEAR{0000}% e.g., 2005
%\FIRSTPAGE{000}%
%\LASTPAGE{000}%
%\SHORTYEAR{00}% shortened year (two-digit)
%\ISSUE{0000} %
%\LONGFIRSTPAGE{0001} %
%\DOI{10.1287/xxxx.0000.0000}%

% Author's names for the running heads
\RUNAUTHOR{Zuk and Kirszenblat}

% Title or shortened title suitable for running heads. Sample:
\RUNTITLE{Ambulance Offload Zone}

% Full title. Sample:
% \TITLE{Bundling Information Goods of Decreasing Value}
% Enter the full title:
\TITLE{A Queueing Model for the Ambulance Ramping Problem with an Offload Zone}

% Block of authors and their affiliations starts here:
% NOTE: Authors with same affiliation, if the order of authors allows,
%   should be entered in ONE field, separated by a comma.
%   \EMAIL field can be repeated if more than one author
\ARTICLEAUTHORS{%
\AUTHOR{Josef Zuk}
\AFF{Defence Science and Technology Group, Melbourne, Australia,
\EMAIL{josef.zuk@defence.gov.au}}
\AUTHOR{David Kirszenblat}
\AFF{Defence Science and Technology Group, Melbourne, Australia,
\EMAIL{david.kirszenblat@defence.gov.au}}
} % end of the block

\ABSTRACT{%
This work develops a methodology for studying the effect of an offload zone on the
ambulance ramping problem using a multi-server, multi-class non-preemptive priority
queueing model that can be treated analytically.
A prototype model for the ambulance/emergency-department interface is constructed,
which is then implemented as a formal discrete event simulation,
and is run as a regenerative steady-state simulation for empirical estimation of
the ambulance queue-length and waiting-time distributions.
The model is also solved by analytical means for explicit and exact representations
of these distributions, which are subsequently tested against simulation results.
A number of measures of performance is extracted, including the mean and 90th percentiles
of the ambulance queue length and waiting time, as well as the average number of
ambulance days lost per month due to offload delay (offload delay rate).
Various easily computable approximations are proposed and tested. In particular,
a closed-form, purely algebraic expression that approximates
the dependence of the offload delay rate on the capacity of the offload zone is proposed.
It can be evaluated directly from model input parameters and is found to be,
for all practical purposes, indistinguishable from the exact result.
}%

% Fill in data. If unknown, outcomment the field
\MSCCLASS{Primary: 90B22; secondary: 60K25, 60J74}
\ORMSCLASS{Primary: Queues: Priority; secondary: Queues: Markovian}
%\HISTORY{This paper was first submitted on April 12, 1922.}

\HISTORY{Date created: October 20, 2023. Last update: January 10, 2024.}

\maketitle
%%%%%%%%%%%%%%%%%%%%%%%%%%%%%%%%%%%%%%%%%%%%%%%%%%%%%%%%%%%%%%%%%%%%%%

% Samples of sectioning (and labeling) in OPRE
% NOTE: (1) \section and \subsection do NOT end with a period
%       (2) \subsubsection and lower need end punctuation
%       (3) capitalization is as shown (title style).
%
%\section{Introduction.}\label{intro} %%1.
%\subsection{Duality and the Classical EOQ Problem.}\label{class-EOQ} %% 1.1.
%\subsection{Outline.}\label{outline1} %% 1.2.
%\subsubsection{Cyclic Schedules for the General Deterministic SMDP.}
%  \label{cyclic-schedules} %% 1.2.1
%\section{Problem Description.}\label{problemdescription} %% 2.

% Main text
\section{Introduction}
\label{Intro}
When a hospital emergency department (ED) is operating at capacity,
incoming ambulance patients cannot be accepted and transferred from the ambulance stretcher
to an ED bed in a timely manner. In this case, paramedics wait with their
patients, either in the ambulance or a hospital corridor,
and continue to provide care until an ED bed becomes available.
This creates what is known as ambulance offload delay (AOD),
and the ensuing queue of ambulances and their crews,
that are unable to return to service and respond to subsequent call-outs,
is called ambulance ramping \citep{NP:AMA23}.
AODs are increasing year-on-year in Australia \citep{NP:AMA21,NP:ABCNews22} and worldwide.
The ambulance ramping problem was particularly severe during the COVID-19 pandemic \citep{NP:Guardian22,NP:Guardian22A}.
The literature on AODs has been reviewed by \citet{NP:Li19}.

The concept of an ambulance offload zone (OZ) is that suitable patients, who have arrived by ambulance
at a time when there are no free places in the hospital ED will,
after an initial triage, be transferred to a facility (the OZ)
where they are looked after by a team of dedicated paramedics until it is possible to place them in the ED.
The Canadian implementation of the OZ is described in \citep{NP:Carter15}.
The Australian equivalent is referred to as the APOT.

The acronym APOT stands for Ambulance Patient Offload Team \citep{NP:AV21}.
We shall also use this term to refer to the facility that handles the patients who
have been offloaded from arriving ambulances.
The size of the APOT will indicate the number of patients that can be simultaneously
looked after in the facility,
rather than the number of paramedics comprising the team.
This paper addresses the methodology to study the ambulance ramping problem within an Australian context
in relation to an APOT (as defined above).
Other studies addressing the ambulance ramping problem have focused on ambulance diversion or allocation
to hospitals with less crowded EDs, which entails a trade-off between longer transportation times
and shorter offload times \citep{NP:Li21}.
We note that the acronym APOT is also commonly used for ambulance-patient offload time,
which refers to the time interval between the arrival of an ambulance patient at a hospital
and the time at which the patient is transferred to an ED bed (or equivalent).

The present work stems from a tentative request made by the Victorian ambulance service to the
Australian Defence Force (ADF) COVID-19 Task Force to carry out a
preliminary analysis on APOT sizing during the COVID-19 pandemic.
The matter was referred to the Defence Science and Technology Group (DSTG),
and was taken up by scientists working within the area of Operations Research.
While ultimately the decision on whether to make the request firm was negative,
in the meantime, the authors of this work had already constructed
a prototype queueing model as a generic concept demonstrator,
and analysed it initially by means of computer discrete event simulation (DES).
It was subsequently realized that analytical approaches,
leading to both numerical and exact representations, were also possible.
The prototype model is based on a non-preemptive Markovian priority queue.
This was considered to be the simplest paradigm that is able to capture
the main relevant characteristics of the operation of the
ambulance/ED interface.
One of the key assumptions in the Australian APOT implementation was that
there would be no COVID-19 patients admitted into the APOT.

% We adopt a queueing theory approach to modelling the ambulance/ED system.
Queueing problems relevant to EDs have been reviewed by \citet{NP:Elalouf22}.
Priority queues in the context of hospital EDs have also been studied by \citet{NP:Hou20},
and in the context of ambulance services by \citet{NP:Taylor80}.
A queueing model to study AODs was previously considered by \citet{NP:Majedi08}.
More recently, \citet{NP:Almehdawe13} considered a Markov queueing network model of multiple EDs,
accounting for both ambulance arrivals and walk-in patients,
and studied the effect of AODs on the performance of a regional emergency medical service.
They used matrix-analytic methods to solve for the steady state probability distributions
of queue lengths and waiting times.
This resulted in a complex and opaque analysis.
In the present study, we confine ourselves to modelling the ambulance queue at a single hospital
and address the question of how much the ambulance queue length and ambulance waiting time
are reduced by the presence of an APOT of a given size.
This enables us to use a simple and explicit method, based on generating functions,
that avoids the need to truncate the problem to finite maximum queue lengths.

The main performance metrics are the queue-length and waiting-time distributions
of the ramped ambulance queue.
Measures of performance (MOPs) include the mean and 90th-percentile of the queue-length and
waiting-time distributions; noting that the median
queue length and waiting time are not generally useful as they are typically zero
due to the discontinuity in the survival functions at the origin.
Another often quoted MOP is the number of ambulance days lost to AOD per month \citep{NP:Li19},
which we shall refer to as the offload delay rate.
We shall compute and plot these as functions of the APOT size.

The main limitation of the present modelling approach is that steady-state behaviour is assumed.
For this reason, we restrict the total traffic intensity $r$ to the ergodic region
\mbox{$0 < r < 1$}.
This is common to all preceding work on the ambulance ramping problem.
It is possible, however, that an overcrowded ED during a pandemic does not attain a steady-state,
and that the total traffic intensity is sufficiently high that the system (as modelled) will
never become empty. In such a case, one would expect the APOT to provide little benefit,
as it will quickly fill up and remain in that state, thus unable to further relieve AOD.
A more likely scenario is that the ED can become overcrowded during peak times, but
the overcrowding is alleviated at subsequent periods of low arrival traffic,
leading to a cyclo-stationary system over a 24-hour period.
This would entail modelling of time-dependent mean arrival rates,
and is beyond the scope of the present study.
We represent a crowded ED by setting the total traffic intensity $r$ to be close to unity.
A DES that replicated the variation in ambulance
arrivals rate throughout the day is reported in \citep{NP:Clarey13}.

The remainder of the paper is organized as follows:
in Section~\ref{QM}, we formulate the queueing model for the problem at hand and describe its
free parameters.
In Section~\ref{DES}, we discuss how to obtain results from the model using DES.
In Section~\ref{QueueLength}, we derive exact expressions for the ambulance queue length based on
an extension of our previous work on non-preemptive Markovian priority queues \citep{NP:Zuk23,NP:Zuk23A,NP:Zuk23C}.
In Section~\ref{WaitTime}, we derive an exact representation of the waiting time for the ambulance queue.
In Section~\ref{ApproxWaitTime}, we propose a simpler approximate form for the ambulance waiting-time distribution.
On the basis of statistical testing against large simulations, we demonstrate in Section~\ref{WaitTest}
that the exact and approximate
distributions are essentially indistinguishable for all practical purposes.
Detailed comparison of theoretical and simulation results is presented in Section~\ref{Results}.
In Section~\ref{Concl}, we offer some concluding remarks,
and identify a few possible avenues for future work.
Various more technical details are relegated to four appendices.

\section{Queueing Model}
\label{QM}
We consider two patient arrival classes into the hospital system, comprising ambulance
arrivals, with a Poisson arrival rate $\lambda_{\text{amb}}$, and walk-ins with a
Poisson arrival rate $\lambda_{\text{wlk}}$.
On arrival, patients undergo an initial triage, at which time they will be
assigned a patient acuity level, determined by the severity of their case,
and this determines their priority for admission to the ED.
Hospitals around the world have differing systems for characterizing patient acuity,
which comprise categorization into various numbers of levels.
In this study, we shall aggregate patient acuity levels into just three: corresponding
to high, intermediate and low priority levels.
Furthermore, we shall assume that all high-priority patients arrive by ambulance, and
that all low-priority arrivals are walk-ins.
We let $\nu_{\text{hi}}$ denote the fraction of all ambulance arrivals that are
high priority, and let $\nu_{\text{lo}}$ denote the fraction of all walks-ins
that are low priority.
Intermediate-priority patients can be both ambulance arrivals and walk-ins.
However, we stipulate that only intermediate-priority ambulance arrivals
are eligible for offload to the APOT. High-priority patients are assumed to
require a higher level of care than is feasible in the APOT.
Moreover, during the Victorian COVID-19 lockdown, it was mandated that
no COVID-19 patients would go to the APOT.
The foregoing assumed characteristics of the priority levels serve to
determine how to aggregate
an arbitrary number of acuity levels into three priorities.

The queueing discipline among the priority levels is taken to be non-preemptive,
as it is reasonable to assume that a patient who has been admitted to the ED
will remain there until treatment is complete.
The queueing discipline within each priority level is assumed to be
first-in, first-out (FIFO).
\citet{NP:Baron19} adopt just two non-preemptive priority levels, corresponding
to acute and non-acute patients.
\citet{NP:Hou20} have previously also considered modelling ED performance where
patient flows with different levels of acuity are formulated based on a priority queueing model,
the queue being modelled as a multi-class, multi-server system with non-preemptive priorities.
In that work, the focus was on reducing waiting times for high-priority patients.
As will also be the case in the present work,
it was assumed that all service times have an exponential
distribution with a common mean, regardless of the priority level involved, and that steady-state
conditions hold.
\citet{NP:Hou20} observe that \emph{`the exact formula is not available to calculate the waiting times
for different priority classes in the general queues due to the complexity of the mathematical structure'},
and proceed to derive the exact mean waiting times for the Markovian case
(previously considered by \citet{NP:Cobham54}).
These form the basis for an approximation for general arrival and service distributions.
The present work shows how to compute, exactly and efficiently, the full distributions for both
waiting times and queue lengths in the Markovian model.

For simplicity, and because we are not attempting to base this methodological
study on real data, we assume a common exponential mean treatment time
\mbox{$\tau_{\text{srv}} \equiv 1/\mu$}
in the ED for all patients.
The number of servers in the Markovian queueing model is denoted $N$, and
is equated with the number of patients that can be simultaneous accommodated
in the ED. We shall refer to this as the number of beds.
The assumptions on arrival and service rates are the same as those adopted by
\citet{NP:Taylor80} in their study of an urban ambulance service
in the framework of a non-preemptive cutoff-priority model with two priority levels.

While ambulance and walk-in patients wait for admission to the ED in different
physical locations, their totality may be considered as a single physical
queue comprising individuals of different priorities.
The physical queue may be decomposed to various virtual queues according to
arrival class, priority level, or both.
While there are two arrival classes
into the hospital system, the ED has three arrival sources,
from the walk-in queue, the ambulance-vehicle queue, and the APOT.
It is useful to make an explicit distinction between the
ambulance-patient queue and the ambulance-vehicle queue.
The ambulance-patient queue comprises all waiting patients who have arrived by
ambulance, which includes those in the APOT.
The ambulance-vehicle queue comprises the ambulances that remain waiting
because their patients have not been offloaded to the APOT or
transferred into the ED.
It is this queue that is the main focus of the present study.

With their definitions clear from the notation, we have partial arrival rates
\begin{equation}
\lambda_{\text{hi}}^{\text{amb}} = \nu_{\text{hi}}\lambda_{\text{amb}} \;, \quad
\lambda_{\text{lo}}^{\text{wlk}} = \nu_{\text{lo}}\lambda_{\text{wlk}} \;.
\end{equation}
It follows immediately that
\begin{equation}
\lambda_{\text{med}}^{\text{amb}} = (1 - \nu_{\text{hi}})\lambda_{\text{amb}} \;, \quad
\lambda_{\text{med}}^{\text{wlk}} = (1 - \nu_{\text{lo}})\lambda_{\text{wlk}} \;.
\end{equation}
In the present model, we have
\mbox{$\lambda_{\text{lo}}^{\text{amb}} = \lambda_{\text{hi}}^{\text{wlk}} = 0$}.
Therefore, the mean arrival rates per priority level are given by
\begin{align}
\begin{aligned}
\lambda_{\text{hi}}  &= \nu_{\text{hi}}\lambda_{\text{amb}} \,, \;
\lambda_{\text{lo}}   = \nu_{\text{lo}}\lambda_{\text{wlk}} \,, \\
\lambda_{\text{med}} &= (1-\nu_{\text{hi}})\lambda_{\text{amb}} + (1-\nu_{\text{lo}})\lambda_{\text{wlk}} \,.
\end{aligned}
\label{PriorityRates}
\end{align}
Accordingly, for the level traffic intensities,
\begin{equation}
r_{\text{lo}}  = \lambda_{\text{lo}}/(N\mu) \,, \;
r_{\text{med}} = \lambda_{\text{med}}/(N\mu) \,, \;
r_{\text{hi}}  = \lambda_{\text{hi}}/(N\mu) \,,
\end{equation}
leading to a total traffic intensity
\mbox{$r = r_{\text{lo}} + r_{\text{med}} + r_{\text{hi}}$}.

For the purposes of model testing, it is convenient to choose a pre-determined total traffic
intensity $r$ in the ergodic range
\mbox{$0 < r < 1$}
at the outset.
Subsequently, the fraction of arrivals by ambulance $\nu_{\text{amb}}$ can be specified.
Thus,
\begin{equation}
r_{\text{amb}} = \nu_{\text{amb}}r \,, \; r_{\text{wlk}} = (1-\nu_{\text{amb}})r \,,
\end{equation}
and
\mbox{$\lambda_{\text{amb}} = N\mu r_{\text{amb}}$},
\mbox{$\lambda_{\text{wlk}} = N\mu r_{\text{wlk}}$}.
Without loss of generality, one can set
\mbox{$\mu = 1$},
which simply means that time is being measured in units of the
common mean service time.
Finally, the APOT size is denoted by $M$.
Therefore, we see that the model is specified entirely by the collection of
dimensionless numbers
\mbox{$\{N,M,r,\nu_{\text{amb}},\nu_{\text{hi}},\nu_{\text{lo}}\}$},
all of which may be chosen independently.
In terms of these parameters, the fractions of all patient arrivals that are
of high, intermediate and low priority are given, respectively, by
\begin{align}
\begin{aligned}
f_{\text{hi}}  &= \nu_{\text{hi}}\nu_{\text{amb}} \,, \\
f_{\text{med}} &= (1-\nu_{\text{hi}})\nu_{\text{amb}} + (1 - \nu_{\text{lo}})(1- \nu_{\text{amb}}) \,, \\
f_{\text{lo}}  &= \nu_{\text{lo}}(1- \nu_{\text{amb}}) \,.
\end{aligned}
\end{align}
In this model, the APOT may be regarded as the first $M$ places in the queue of intermediate-level
ambulance-patient arrivals.

In Victoria, a set of five triage categories is used, as summarized in Table~\ref{tab:data}.
The triage category indicates the urgency of the patient’s need for medical care.
We exclude the most urgent category T1 from the present modelling,
as these patients require immediate attention
and are assumed to be dealt with outside the ED system
via a separate dedicated trauma facility.
The high-priority patient level is equated with T2.
The low-priority patient level is equated with T5, and the
intermediate-priority patient level is taken to be the aggregation of T3 and T4.
Statewide data for Victoria on the number of presentations to EDs during the
July--September quarter of 2022 in each triage category \citep{NP:VAHI}
is presented in Table~\ref{tab:data},
as is the number of emergency patients transported to the ED during this period.
Given these data, one finds that
\mbox{$\nu_{\text{amb}} = 0.248$},
\mbox{$f_{\text{hi}} = 0.163$}
and
\mbox{$f_{\text{lo}} = 0.060$},
from which it follows that
\mbox{$\nu_{\text{hi}} = 0.66$},
\mbox{$\nu_{\text{lo}} = 0.08$}.
For our standard test case, we adopt the parameter values quoted in Table~\ref{tab:parm}.
We adopt a larger value for $\nu_{\text{amb}}$ than the data suggests in order to represent
a more extreme (pandemic-like) scenario.

\begin{table}
\TABLE
%\centering
{Emergency Patient Data: Jul--Sep 2022\label{tab:data}}
{\begin{tabular}{|l|l|l|c|}
\hline
Category & Name & Description & Count \\
\hline\hline
T1 & Resuscitation & requires treatment immediately       & 3663 \\
T2 & Emergency     & requires treatment within 10 minutes & 75,170 \\
T3 & Urgent        & requires treatment within 30 minutes & 197,170 \\
T4 & Semi-Urgent   & requires treatment within 1 hour     & 160,600 \\
T5 & Non-Urgent    & requires treatment within 2 hours    & 27,573 \\
\hline
Amb & \multicolumn{2}{l|}{Emergency patients transported by ambulance} & 118,056 \\
\hline
\end{tabular}}
{}
\end{table}

\begin{table}
\TABLE
%\centering
{Model Parameters\label{tab:parm}}
{\begin{tabular}{|l|c|c|}
\hline
Description & Symbol & Value\\
\hline\hline
ED size (number of servers)                           & $N$                & $10$   \\
APOT size                                             & $M$                & $6$    \\
Mean treatment time                                   & $1/\mu$            & $1$    \\
Total traffic intensity                               & $r$                & $0.95$  \\
Fraction of ambulance arrivals                        & $\nu_{\text{amb}}$ & $2/3$  \\
Fraction of ambulance arrivals that are high priority & $\nu_{\text{hi}}$  & $2/3$ \\
Fraction of walk-in arrivals that are low priority    & $\nu_{\text{lo}}$  & $0.10$ \\
\hline
\end{tabular}}
{}
\end{table}

The purpose of the present paper is to develop computational techniques,
not to analyse real data, nor make recommendations about actual systems.
Therefore, the model studied here
is a prototype that serves as a concept demonstrator for a simple queueing approach
to the problem, and is chosen to be the simplest exponent that accounts for the
main characteristics of the ambulance/ED interface relevant to the ramping issue.
These include the following:
\begin{itemize}
\item
ambulance arrivals compete with walk-ins for admission to the ED;
\item
patients have different priorities for admission to the ED;
\item
both ambulance arrivals and walk-ins are a mixture of priorities;
\item
not all ambulance arrivals are eligible for offload to the APOT.
\end{itemize}

\section{Discrete Event Simulation}
\label{DES}
We implemented the ambulance queueing model in {\sc Matlab} as formal DES.
There are three events that may be scheduled on the event list as indicated in Table~\ref{tab:enums}.
The entries on the event list are pairs comprising an epoch and the index of an event type.
The simulation is started by initializing the event list with a single \texttt{AmbulanceArrival} event
and a single \texttt{WalkInArrival} event at epochs corresponding to the relevant exponentially distributed
time intervals after epoch zero.
Each \texttt{AmbulanceArrival} event schedules the next one.
Each \texttt{WalkInArrival} event schedules the next one.

The process associated with the \texttt{AmbulanceArrival} event randomly assigns
the patient's priority level. If there are no idle servers, it adds the patient
to the ambulance queue or the APOT. Otherwise, it calls the \texttt{PatientTreatment} process.
The process associated with the \texttt{AmbulanceArrival} event is similar, expect
that it adds the patient to the walk-in queue if there are no idle servers.
The \texttt{PatientTreatment} process randomly assigns a treatment time, adds an entry to the
\texttt{PatientHistory} data structure, and schedules an \texttt{EndService} event.
The process associated with the \texttt{EndService} event selects the next patient to be treated
from the appropriate queue, and calls the \texttt{PatientTreatment} process. If a patient from
the APOT is selected, then it also calls the
\texttt{APOTTransfer} process that manages the transfer of a patient from the ambulance queue
to the APOT when an APOT place becomes free.

The sole output of the simulation is the patient history, which is an array whose
elements are data structures having the following fields:
\vspace{0.25\baselineskip}
\nextverbatimspread{1}
\begin{verbatim}
PatientHistory.ArrivalTime
PatientHistory.WaitTime
PatientHistory.TreatmentTime
PatientHistory.APOTTime
PatientHistory.Source
PatientHistory.Level
\end{verbatim}
\vspace{0.25\baselineskip}
There is one element for each patient processed in the simulation.
The \texttt{ArrivalTime} field is the epoch at which a patient arrives in the system.
The \texttt{WaitTime} field is the time that the patient waits before entering the ED.
The \texttt{TreatmentTime} field is the time that the patient spends in the ED.
The \texttt{APOTTime} field is the time that the patient spends in the APOT.
The \texttt{Source} field records from where the patient arrived into the ED as indicated in Table~\ref{tab:enums}.
The \texttt{Level} field is the patient acuity level, which is equated with the priority level, as indicated in Table~\ref{tab:enums}.

\begin{table}
\TABLE
%\centering
{Enumerations\label{tab:enums}}
{\begin{tabular}{|c|l|l|l|}
\hline
Index & Events & Sources & Levels \\
\hline\hline
$1$ & \texttt{AmbulanceArrival} & \texttt{Ambulance} & \texttt{Lo}  \\
$2$ & \texttt{WalkInArrival}    & \texttt{WalkIn}    & \texttt{Med} \\
$3$ & \texttt{EndService}       & \texttt{APOT}     & \texttt{Hi}  \\
\hline
\end{tabular}}
{}
\end{table}

All quantities of interest can be extracted for the patient history.
In particular, we are able to plot the marginal queue-length and waiting-time distributions for:
\begin{enumerate}
\item
All high, intermediate and low priority patients.
\item
The aggregation of all patients.
\item
All ambulance and walk-in patients.
\item
All combinations of arrival class and priority level.
\item
The APOT.
\item
The ambulance-vehicle queue.
\end{enumerate}
Item~6 is the main object of the study. Items~1-2 serve as useful checks of the
simulation as they can be compared with previously established theoretical results.
Item~1 can be compared against the explicit analytical results on the non-preemptive
Markovian priority queue derived in \citep{NP:Zuk23C,NP:Zuk23}.
Item~2 can be compared with the expected geometric and exponential distributions.
Analytic results for item~6 are derived in this paper, and the concommitant
results for items~3--5 are generated as by-products.
The partial busy period distribution is also constructed from the patient history and
can be compared with the theoretical results derived in \citep{NP:Zuk23B}.

\section{Ambulance Queue Length}
\label{QueueLength}
We begin by considering the full three-dimensional joint queue-length probability mass function (PMF)
\mbox{$P_{\text{L3}}(\ell,m,n)$}
that gives the probability of finding exactly $\ell$ high-priority, $m$ intermediate-priority
and $n$ low-priority patients in the queue, regardless of their arrival class.
The wait-conditional three-dimensional joint queue-length PMF
\mbox{$P_{\text{L3}}^{\text{c}}(\ell,m,n)$}
is obtained from
\begin{equation}
P_{\text{L3}}(\ell,m,n) = P_{\text{NW}}\delta_{\ell 0}\delta_{n0}\delta_{m0}
     + (1 - P_{\text{NW}})P_{\text{L3}}^{\text{c}}(\ell,m,n) \;.
\label{PFull}
\end{equation}
This PMF gives the probability of finding the indicated number of patients in the queue,
in steady state, conditional on the ED being full.
It is also referred to as the delay-conditional PMF \citep{NP:Davis66}.
The no-wait probability $P_{\text{NW}}$
corresponds to the probability that a new arrival will find at least one free place in the ED.
It is independent of the queue discipline, and
its value is given by \citep{NP:Davis66}
\begin{equation}
\frac{1}{1 - P_{\text{NW}}} = 1 + (1-r)\frac{N!}{(Nr)^N}{\cdot}
     \sum_{k=0}^{N-1} \frac{(Nr)^k}{k!} \;.
\label{PNW}
\end{equation}
Since our focus is on the ambulance-vehicle distribution, and there are assumed to be no
ambulance patients of low priority,
the low-priority level is irrelevant and can be marginalized out.
This leads to consideration of the two-dimensional joint queue-length PMF
\begin{equation}
P_{\text{L2}}^{\text{c}}(\ell,m) = \sum_{n=0}^\infty P_{\text{L3}}^{\text{c}}(\ell,m,n) \;,
\end{equation}
that gives the probability of finding exactly $\ell$ high-priority and $m$ intermediate-priority
patients in the queue, regardless of their arrival class.
This is the PMF for a two-level non-preemptive priority model having a lower priority level
with mean arrival rate $\lambda_{\text{med}}$ and a higher priority level with
mean arrival rate $\lambda_{\text{hi}}$,
as given by (\ref{PriorityRates}).
The total traffic intensity within this reduced model is given by
\mbox{$\sigma \equiv r_{\text{med}} + r_{\text{hi}}$}.
The dependence of the model on the low-priority level traffic intensity arises entirely
when the full model is recovered via the analogue of (\ref{PFull}),
which involves the overall total traffic intensity $r$.
The representation and computation of the joint queue-length PMFs
\mbox{$P_{\text{L3}}^{\text{c}}(\ell,m,n)$},
\mbox{$P_{\text{L2}}^{\text{c}}(\ell,m)$}
for the purely non-preemptive priority queueing model has been discussed extensively
in \citep{NP:Zuk23C,NP:Zuk23,NP:Zuk23A}.
Let $p$ denote the fraction of intermediate-level patients that arrive by ambulance and
\mbox{$q \equiv 1 - p$},
so that $q$ represents the fraction of intermediate-level patients that are walk-ins. Then,
\begin{equation}
p = \frac{\lambda_{\text{med}}^{\text{amb}}}{\lambda_{\text{med}}^{\text{amb}} + \lambda_{\text{med}}^{\text{wlk}}}
\;, \quad
q = \frac{\lambda_{\text{med}}^{\text{wlk}}}{\lambda_{\text{med}}^{\text{amb}} + \lambda_{\text{med}}^{\text{wlk}}} \;,
\label{pq}
\end{equation}
in terms of which the ambulance-patient queue-length PMF can be expressed as
\begin{equation}
P_{\text{amb}}^{\text{c}}(\ell,m) = \sum_{n=m}^\infty P_{\text{L2}}^{\text{c}}(\ell,n)
     {\cdot}\binom{n}{m} p^m q^{n-m} \;.
\end{equation}
The infinite summation can be avoided.
Let $G_\ell^{\text{c}}(z)$ be the joint probability generating function (PGF) associated with the two-dimensional PMF
$P_{\text{L2}}^{\text{c}}(\ell,m)$,
such that
\begin{equation}
G_\ell^{\text{c}}(z) \equiv \sum_{m=0}^\infty P_{\text{L2}}^{\text{c}}(\ell,m)z^m \;, \quad
     P_{\text{L2}}^{\text{c}}(\ell,m) = \oint_\mathcal{C} \frac{dz}{2\pi i}\,
     \frac{1}{z^{m+1}}G_\ell^{\text{c}}(z) \;,
\label{PGFC}
\end{equation}
where the contour $\mathcal{C}$ is an infinitesimal anti-clockwise circle around the origin.
The PGF $G_\ell(z)$ for the unconditional distribution is given by
\begin{equation}
G_\ell(z) = P_{\text{NW}}\delta_{\ell 0} + (1 - P_{\text{NW}})G_\ell^{\text{c}}(z) \;,
\end{equation}
where $P_{\text{NW}}$ denotes the no-wait probability.
In view of (\ref{PGFC}), we have
\begin{align}
\begin{aligned}
P_{\text{amb}}^{\text{c}}(\ell,m) &= (p/q)^m\sum_{n=m}^\infty \binom{n}{m}
     q^n\oint_\mathcal{C} \frac{dz}{2\pi i}\, \frac{1}{z^{n+1}}G_\ell^{\text{c}}(z) \\
&= (p/q)^m\oint_\mathcal{C} \frac{dz}{2\pi iz}\, G_\ell^{\text{c}}(z)
     \sum_{n=m}^\infty \binom{n}{m} (q/z)^n \\
&= p^m\oint_\mathcal{C} \frac{dz}{2\pi i}\, \frac{1}{z^{m+1}}G_\ell^{\text{c}}(z+q) \;,
\end{aligned}
\end{align}
where we have made use of the result
%%% NB: This equation could be omitted.
\begin{equation}
f_k(z) \equiv \sum_{n=k}^\infty \binom{n}{k} z^n = \frac{z^k}{(1-z)^{k+1}} \;.
\end{equation}
We see from this that the PGF associated with
\mbox{$P_{\text{amb}}^{\text{c}}(\ell,m)$}
is
\mbox{$G_\ell^{\text{c}}(pz+q)$}.

It has been established in \citep{NP:Cohen56} and \citep{NP:Zuk23C} that
\begin{equation}
G^{\text{c}}_\ell(z) = \frac{1-\rr}{1 - \rr z}{\cdot}\left[1 - z\zeta_-(z)\right]\zeta_-^\ell(z) \;,
\end{equation}
where the quantities $\zeta_\pm(z)$ pertain to the low-priority marginal PGF,
and are discussed in Appendix~\ref{MargPGF}.
Hence,
\begin{align}
\begin{aligned}
P_{\text{amb}}^{\text{c}}(\ell,m) &= p^m\oint_\mathcal{C}\frac{dz}{2\pi i}\,
     \frac{1}{(z-q)^{m+1}}{\cdot}G^{\text{c}}_\ell(z) \\
&= p^m\oint_\mathcal{C}\frac{dz}{2\pi i}\, \frac{1-\rr}{1-\rr z}{\cdot}
     \frac{\zeta_-^\ell(z)}{(z-q)^{m+1}}
%\\ &\quad {}
     - p^m\oint_\mathcal{C}\frac{dz}{2\pi i}\, \frac{1-\rr}{1-\rr z}{\cdot}
     \frac{\zeta_-^{\ell+1}(z)}{(z-q)^{m}}
\\ &\quad {}
     - qp^m\oint_\mathcal{C}\frac{dz}{2\pi i}\, \frac{1-\rr}{1-\rr z}{\cdot}
      \frac{\zeta_-^{\ell+1}(z)}{(z-q)^{m+1}} \;,
\end{aligned}
\end{align}
where the contour $\mathcal{C}$ is now centred on
\mbox{$z = q$}.
The integration contour can be deformed \citep{NP:Zuk23C} into one that encircles a pole at
\mbox{$x_{\text{pol}} = 1/r$}
on the real axis, and a cut along the real interval
\mbox{$[x_-,x_+]$},
with the definitions
\begin{equation}
x_\pm \equiv 1 + (1 \pm \sqrt{r_{\text{hi}}})^2/r_{\text{med}} \,,
     \; x_{\text{dif}} \equiv x _+ - x_- \,.
\end{equation}
We also introduce
\begin{equation}
a_q \equiv (x_- - q)/x_{\text{dif}} \,, \;
     b \equiv (x_- - 1/\rr)/x_{\text{dif}} \,.
\end{equation}
In terms of the basic model parameters, we can write
\begin{equation}
a_q = (1 + p\rr + q r_{\text{hi}})/(4\sqrt{r_{\text{hi}}}) - 1/2 \,, \;
b   = (\rr/\sqrt{r_{\text{hi}}} + \sqrt{r_{\text{hi}}}/\rr)/4 - 1/2 \;.
\end{equation}

In the pole/cut decomposition
%%% NB: We could inline this equation.
%\begin{equation}
\mbox{$P_{\text{amb}}^{\text{c}}(\ell,m) = P_{\text{pol}}^{\text{c}}(\ell,m) + P_{\text{cut}}^{\text{c}}(\ell,m)$},
%\end{equation}
the cut contribution is given by
\begin{equation}
\begin{split}
P_{\text{cut}}^{\text{c}}(\ell,m) &= p^m\Lambda_q(\ell+1,m) - p^m\Lambda_q(\ell,m+1)
%\\ &\quad {}
     + qp^m\Lambda_q(\ell+1,m+1) \;,
\end{split}
\end{equation}
where
\begin{equation}
\Lambda_q(\ell,m) \equiv \frac{2(1-\rr)}{\rr}{\cdot}\frac{r_{\text{hi}}^{\ell/2}}{x_{\text{dif}}^m}
     \int_0^1 d\tau \frac{[1 - u(\tau)]C_\ell(u(\tau))}{[a_q + u(\tau)]^m[1 + b/u(\tau)]} \,,
\end{equation}
with
\mbox{$u(\tau) = \cos^2(\pi\tau/2)$}
and
\begin{equation}
C_\ell(u) \equiv \frac{\sin(2\ell\asin\sqrt{u})}{\sin(2\asin\sqrt{u})}
     = (-1)^{\ell-1}\frac{\sin(\ell\pi\tau)}{\sin(\pi\tau)} \,,
     \quad C_\ell(0) = \ell \,.
\end{equation}
Following the treatment in \citep{NP:Zuk23C},
an $L$-point Gaussian quadrature rule reads
\begin{equation}
\Lambda_q(\ell,m) \simeq r_{\text{hi}}^{\ell/2}\sum_{k=1}^L\frac{W_kC_\ell(U_k)}
     {[x_{\text{dif}}(a_q + U_k)]^m} \;,
\label{LamQuad}
\end{equation}
where the quadrature nodes and weights $U_k$, $W_k$, respectively, are given by
\begin{equation}
U_k \equiv \cos^2(\pi\tau_k/2)\,, \;
W_k \equiv \frac{2(1-\rr)}{\rr L}{\cdot}\frac{1 - U_k}{1 + b/U_k} \,.
\label{UWGQ}
\end{equation}
The choice of integration grid
\mbox{$\tau_k = k/L$},
\mbox{$k = 1,2,\ldots,L$},
corresponds to a Gauss-Chebyshev quadrature of the second kind, implemented as an
exponentially convergent trapezoidal rule.
The choice of integration grid
\mbox{$\tau_k = (k-1/2)/L$},
\mbox{$k = 1,2,\ldots,L$},
corresponds to a Gauss-Chebyshev quadrature of the first kind, implemented as an
exponentially convergent mid-point rule.
Both rules can be applied iteratively (increasing $L$ at each step) to
attain a prescribed level of accuracy.
The pole contribution is given by
\begin{equation}
P_{\text{pol}}^{\text{c}}(\ell,m) =
     \frac{(1-\rr)(p\rr)^m}{(1 - q\rr)^{m+1}}\left(1 - \frac{r_{\text{hi}}}{\rr^2}\right){\cdot}
     \left(\frac{r_{\text{hi}}}{\rr}\right)^\ell\!\!{\cdot} \Theta(\rr^2 - r_{\text{hi}}) \;.
\label{pjointpol}
\end{equation}

We want to derive a PMF for just the ambulance-vehicle queue, which is the subset of the
ambulance-patient queue excluding the APOT. To do so,
we account for the presence of an APOT of size $M$ by introducing a shift
in the intermediate-priority queue-length variable, according to
\begin{equation}
P^{\text{c}}_M(\ell,n) = \left\{\begin{array}{ccc}
\sum_{m=0}^M P_{\text{amb}}^{\text{c}}(\ell,m) & \text{for} & n = 0 \;, \\
P_{\text{amb}}^{\text{c}}(\ell,n+M)            & \text{for} & n \geq 1 \;,
\end{array}
\right.
\end{equation}
which may equivalently be expressed as
\begin{equation}
P^{\text{c}}_M(\ell,n) = \delta_{n0}{\cdot}\!\!\sum_{m=0}^{M-1} P_{\text{amb}}^{\text{c}}(\ell,m)
     + P_{\text{amb}}^{\text{c}}(\ell, n+M) \;.
\label{PM}
\end{equation}
For the aggregated ambulance-vehicle queue,
comprising both high and intermediate level patients,
we have the convolution
\begin{equation}
P_{\text{amb}}^{\text{c}}(n) = \sum_{m = 0}^n P_M^{\text{c}}(n-m,m)  \;.
\end{equation}
Marginal PMFs for the single-class multi-level non-preemptive Markovian priority queue
could be derived from the results on the waiting-time marginals presented in \citep{NP:Davis66}
by invoking the distributional form of Little's law, as discussed in \cite{NP:Zuk23C}.
However, these are insufficient for the calculation of $P_{\text{amb}}^{\text{c}}(n)$,
where the knowledge of the joint level PMF is required,
and which is only made possible due to the explicit and computationally efficient
representations of the joint level PMF
recently developed in \citep{NP:Zuk23C,NP:Zuk23,NP:Zuk23A}.

We have the result
\begin{align}
\begin{aligned}
G^{(M)}_\ell(z) &\equiv \sum_{n=0}^\infty P^{\text{c}}_M(\ell,n) z^n
%\\ &
     = \frac{G^{\text{c}}_\ell(z)}{z^M} + \sum_{m=0}^{M-1} P_{\text{amb}}^{\text{c}}(\ell,m)
     \left(1 - \frac{1}{z^{M-m}}\right) \,.
\end{aligned}
\end{align}
For the intermediate-level marginal, this reduces to
\begin{align}
\begin{aligned}
g^{(M)}_{\text{med}}(z) &\equiv \sum_{\ell=0}^\infty G^{(M)}_\ell(z)
%\\ &
     = \frac{g_{\text{med}}(z)}{z^M} + \sum_{m=0}^{M-1} P_{\text{med}}^{\text{c}}(m)
     \left(1 - \frac{1}{z^{M-m}}\right) \,,
\end{aligned}
\end{align}
where
\mbox{$P^{\text{c}}_{\text{med}}(n) \equiv \sum_{\ell=0}^\infty P_{\text{amb}}^{\text{c}}(\ell,n)$}
is the queue-length marginal for all intermediate ambulance arrivals (including those in the APOT).
The (wait-conditional) mean queue-length of intermediate-level patients in the ambulance-vehicle
in the presence of an
APOT of size $M$ then follows directly from
the derivative at
\mbox{$z = 1$},
\mbox{$\overline{L}\vphantom{L}^{\text{c}}_M = g^{(M)\prime}_{\text{med}}(1)$},
and is given by
\begin{equation}
\overline{L}\vphantom{L}^{\text{c}}_M \equiv \sum_{n=0}^\infty nP^{\text{c}}_M(n)
    = \overline{L}\vphantom{L}^{\text{c}}_{\text{med}} - M + \sum_{m=0}^{M-1} (M-m)P_{\text{med}}^{\text{c}}(m) \;,
\label{LM}
\end{equation}
where
\mbox{$\overline{L}\vphantom{L}^{\text{c}}_{\text{med}}$}
is the (wait-conditional) mean queue-length
of all intermediate-level ambulance patients (including those in the APOT).
It is given by
\mbox{$\overline{L}\vphantom{L}^{\text{c}}_{\text{med}} = p\overline{L}\vphantom{L}^{\text{c}}_{2}$}
where
\mbox{$\overline{L}\vphantom{L}^{\text{c}}_{2}$}
represents the (wait-conditional) mean queue-length
of all intermediate-level patients (ambulance plus walk-ins)
as derived in Appendix~\ref{Means}
and given by (\ref{LWmed}).
The (wait-conditional) mean length of the entire ambulance vehicle queue in the presence of an APOT is given by
\mbox{$\overline{L}\vphantom{L}^{\text{c}}_{\text{amb}} = \overline{L}\vphantom{L}^{\text{c}}_{\text{hi}}
    + \overline{L}\vphantom{L}^{\text{c}}_M$},
where
\mbox{$\overline{L}\vphantom{L}^{\text{c}}_{\text{hi}} = \overline{L}\vphantom{L}^{\text{c}}_{1}$}
is the (wait-conditional) mean queue-length
of the high-level ambulance patients, as derived in Appendix~\ref{Means}
and given by (\ref{LWhi}).

The wait-conditional PMF for APOT occupancy is given by
\begin{equation}
P^{\text{c}}_{\text{apot}}(m) = \left\{
\begin{array}{ccl}
P^{\text{c}}_{\text{med}}(m)                    & \quad\text{for}\quad & m = 0,1,\ldots,M-1 \,, \\
1 - \sum_{j=0}^{M-1}P^{\text{c}}_{\text{med}}(j) & \quad\text{for}\quad & m = M  \,.
\end{array}
\right.
\end{equation}

\section{Ambulance Waiting Time}
\label{WaitTime}
Let us consider the intermediate-level marginal for the ambulance-vehicle queue length
\begin{equation}
P^{\text{c}}_M(n) \equiv \sum_{\ell=0}^\infty P^{\text{c}}_M(\ell,n)
     = \delta_{n0}\sum_{m=0}^{M-1} P^{\text{c}}_{\text{med}}(m) + P^{\text{c}}_{\text{med}}(n+M) \;,
\label{PMambv}
\end{equation}
where
\mbox{$P^{\text{c}}_{\text{med}}(m) \equiv \sum_{\ell=0}^\infty P_{\text{amb}}^{\text{c}}(\ell,m)$}
is the intermediate-level marginal of the ambulance-patient queue.
Let $\chi$ denote the probability that the APOT is full. Then,
\begin{equation}
\chi = \sum_{m = M}^\infty P_{\text{med}}^{\text{c}}(m) = 1 -
     \sum_{m = 0}^{M-1} P_{\text{med}}^{\text{c}}(m) \;.
\label{ChiDef}
\end{equation}
Thus, we can rewrite (\ref{PMambv}) as
\begin{equation}
P^{\text{c}}_M(n) = (1-\chi)\delta_{n0} + \chi P^{\text{c}\#}_M(n) \;,
\end{equation}
where
\mbox{$P^{\text{c}\#}_M(n) \equiv P^{\text{c}}_{\text{med}}(n+M)/\chi$}
represents the probability that there are $n$ patients in the ambulance-vehicle
queue conditional on the APOT being full.
The PGF associated with the PMF
\mbox{$P^{\text{c}\#}_M(n)$}
is given by
\begin{equation}
g^{\#}_M(z) \equiv \sum_{n=0}^\infty P^{\text{c}\#}_M(n) z^n
     = \frac{g_{\text{med}}(z)}{\chi z^M} - \frac{1}{\chi}\sum_{m=0}^{M-1}
     \frac{P_{\text{med}}^{c}(n)}{z^{M-n}} \;,
\end{equation}
where
\begin{equation}
g_{\text{med}}(z) = g(pz+q) \,, \; g(z) \equiv \sum_{\ell=0}^\infty G^{\text{c}}_\ell(z) \;.
\end{equation}
We note that that second term on the RHS is analytic in the complex $z$ plane
apart from a pole at the origin.
The PGF associated with
\mbox{$P^{\text{c}}_M(n)$}
is given by
\begin{equation}
g_M(z) \equiv \sum_{n=0}^\infty P^{\text{c}}_M(n) z^n
     = 1 - \chi + \chi g^{\#}_M(z) \;.
\end{equation}

Let
\mbox{$\bar{F}^{(M)}_{\text{med}}(t)$}
denote the (wait-conditional) survival function (SF) of the waiting time for the
intermediate-level ambulance patients in the presence of an APOT of size $M$.
Knowledge of $g^{\#}_M(z)$ allows us to compute $\bar{F}^{(M)}_{\text{med}}(t)$
via the distributional form of Little's law.
The derivation is presented in Appendix~\ref{Wait}.

Let
\mbox{$\bar{\nu}_{\text{hi}} \equiv 1 - \nu_{\text{hi}}$},
which represents the fraction of ambulance-arrivals that are not high priority
({\it i.e.}\ intermediate priority).
The fraction of all patients arriving by ambulance that will enter the
ambulance-vehicle queue is given by
\mbox{$1 - (1-\chi)\bar{\nu}_{\text{hi}} = \nu_{\text{hi}} + \chi\bar{\nu}_{\text{hi}}$},
as only intermediate-level ambulance arrivals will avoid the queue, provided that the
APOT is not full.
This fraction is partitioned as the sum of the ratios
\mbox{$\nu_{\text{hi}}/(\nu_{\text{hi}} + \chi\bar{\nu}_{\text{hi}})$}
of the high-priority patients, and
\mbox{$\chi\bar{\nu}_{\text{hi}}/(\nu_{\text{hi}} + \chi\bar{\nu}_{\text{hi}})$}
of the intermediate-priority patients.
The waiting time for the high-priority patients is exponentially distributed,
and the wait-conditional SF of the waiting-time in the ambulance-vehicle queue
is given by the mixture
\begin{equation}
\bar{F}^{\text{c}}_{\text{amb}}(t) = \frac{\nu_{\text{hi}}}{\nu_{\text{hi}} + \chi\bar{\nu}_{\text{hi}}}
     e^{-(1-r_{\text{hi}})N\mu t} + \frac{\chi\bar{\nu}_{\text{hi}}}{\nu_{\text{hi}} + \chi\bar{\nu}_{\text{hi}}}
     \bar{F}^{(M)}_{\text{med}}(t) \;.
\label{SFWaitAmb}
\end{equation}
The unconditional waiting-time distribution is recovered via
\begin{align}
\begin{aligned}
P_{\text{amb}}(t)       &= P_{\text{NW}}\delta(t) + (1 - P_{\text{NW}})P^{\text{c}}_{\text{amb}}(t) \;, \\
\bar{F}_{\text{amb}}(t) &= P_{\text{NW}}[1 - \Theta(t)] + (1 - P_{\text{NW}})\bar{F}^{\text{c}}_{\text{amb}}(t) \;,
\end{aligned}
\end{align}
for the probability density function (PDF) and SF, respectively, or
\begin{equation}
\bar{F}_{\text{amb}}(t) = \left\{\begin{array}{ccc}
1                                                     & \text{for} & t = 0 \;, \\
(1 - P_{\text{NW}})\bar{F}^{\text{c}}_{\text{amb}}(t) & \text{for} & t > 0 \;.
\end{array}
\right.
\end{equation}
This is the exact result for the quantity that we shall compare with simulation.

\subsection{Approximate Waiting Time}
\label{ApproxWaitTime}
A simple approximation for the waiting-time distribution of the ambulance-vehicle queue
can be obtained from the ansatz that it is a mixture of the marginal PDFs for the
aggregated high-priority waiting-time distribution and the aggregated
intermediate-priority waiting-time distribution.
Thus, we write for the wait-conditional PDF and SF, respectively,
\begin{align}
\begin{aligned}
P^{\text{c}}_{\text{amb}}(t)       &= \alpha P^{\text{c}}_{1}(t) + (1-\alpha)P^{\text{c}}_{2}(t) \,, \\
\bar{F}^{\text{c}}_{\text{amb}}(t) &= \alpha \bar{F}^{\text{c}}_{1}(t) + (1-\alpha)\bar{F}^{\text{c}}_{2}(t) \,.
\end{aligned}
\end{align}
Here,
\mbox{$P^{\text{c}}_{1}(t)$},
\mbox{$P^{\text{c}}_{2}(t)$}
are the marginal waiting-time PDFs for the high and low priority levels, respectively,
of the generic two-level non-preemptive priority queue.
In the present application, the high-level traffic intensity is
\mbox{$r_1 = r_{\text{hi}}$}
and the low-level traffic intensity is
\mbox{$r_2 = r_{\text{med}}$}.
The total traffic intensity is the sum of these.
The high-priority marginal distribution is exponential:
\begin{equation}
P^{\text{c}}_{1}(t) = (1-r_1)e^{-(1-r_1)t} \;,
\end{equation}
while the low-priority marginal distribution is related to the Humbert
bivariate confluent hypergeometric function
\mbox{$\Phi_1(\alpha,\beta,\gamma;x,y)$} \citep{NP:Humbert20,NP:Humbert22},
and has integral representation
\begin{align}
\begin{aligned}
P^{\text{c}}_{2}(t) &= -(1-r_{\text{sum}})\left(\frac{r_{1}}{r_{\text{sum}}^2}-1\right)
     \Theta(r_{\text{sum}}^2-r_{1})
     e^{-tr_{2}(1/r_{\text{sum}}-1)} \\
     &\quad {}+ \frac{2(1-r_{\text{sum}})\sqrt{r_{1}}}{\pi r_{\text{sum}}}e^{-t(4\sqrt{r_{1}}a - r_{2})}
     \int_0^1 du\, e^{-4t\sqrt{r_{1}}{\cdot}u}{\cdot}\frac{\sqrt{u(1-u)}}{u+b} \;,
\end{aligned}
\label{WaitPDF}
\end{align}
with
\mbox{$r_{\text{sum}} \equiv r_1+r_2$}
and
\begin{equation}
a \equiv \frac{1+r_{\text{sum}}}{4\sqrt{r_{1}}} - \frac{1}{2}\;, \quad
     b \equiv \frac{1}{4}\left(\frac{r_{\text{sum}}}{\sqrt{r_{1}}} +
     \frac{\sqrt{r_{1}}}{r_{\text{sum}}}\right) - \frac{1}{2}\;.
\end{equation}
The derivation of these results, and the efficient computation of
\mbox{$P^{\text{c}}_{2}(t)$}
are discussed in \citep{NP:Zuk23C}.
The initial discussion of an integral representation for waiting time distributions
in a non-preemptive priority queue was provided by \citet{NP:Davis66}.

For the mean waiting time, we have
\begin{equation}
\overline{W}\vphantom{W}^{\text{c}}_{\text{amb}} = \alpha\overline{W}\vphantom{W}^{\text{c}}_{1}
     + (1-\alpha)\overline{W}\vphantom{W}^{\text{c}}_{2} \;,
\end{equation}
so that
\begin{equation}
\alpha = \frac{\overline{W}\vphantom{W}^{\text{c}}_{\text{amb}} - \overline{W}\vphantom{W}^{\text{c}}_{2}}
     {\overline{W}\vphantom{W}^{\text{c}}_{1} - \overline{W}\vphantom{W}^{\text{c}}_{2}} \;.
\end{equation}
The level means $\overline{W}\vphantom{W}^{\text{c}}_{1}$, $\overline{W}\vphantom{W}^{\text{c}}_{2}$
follow from (\ref{LWmed}) and (\ref{LWhi}) of Appendix~\ref{Means}, and are given by
\mbox{$\overline{W}\vphantom{W}^{\text{c}}_{1} = 1/(1-r_1)$},
\mbox{$\overline{W}\vphantom{W}^{\text{c}}_{2} = \overline{W}\vphantom{W}^{\text{c}}_{1}/(1-r_{\text{sum}})$},
noting that we set $N\mu = 1$ here.

The coefficient $\alpha$ can be estimated by appealing to Little's law for the mean
ambulance-vehicle waiting time.
By analogy with (\ref{SFWaitAmb}), we set
\mbox{$\alpha = \nu_{\text{hi}}/(\nu_{\text{hi}}+\bar{\nu}_{\text{hi}}\chi)$},
in which case
\mbox{$\chi = (1/\alpha-1)\nu_{\text{hi}}/\bar{\nu}_{\text{hi}}$}.
Here, the quantity $\chi$ is to be determined, rather than set to the probability
of the APOT being full, as we did previously in (\ref{ChiDef}).
We can write it as
\mbox{$\chi = \beta \nu_{\text{hi}}/\bar{\nu}_{\text{hi}}$},
where
\begin{equation}
\beta = \frac{1}{\alpha} - 1 =
     \frac{\overline{W}\vphantom{W}^{\text{c}}_{1} - \overline{W}\vphantom{W}^{\text{c}}_{\text{amb}}}
     {\overline{W}\vphantom{W}^{\text{c}}_{\text{amb}} - \overline{W}\vphantom{W}^{\text{c}}_{2}} \;,
\end{equation}
and compute the ambulance-vehicle mean via
\mbox{$\overline{W}\vphantom{W}^{\text{c}}_{\text{amb}} = \overline{L}\vphantom{W}^{\text{c}}_{\text{amb}}/r_{\text{amb}}$}
with
\mbox{$\overline{L}\vphantom{W}^{\text{c}}_{\text{amb}}$}
as given at the end of the preceding section.
The optimal value of $\beta$ can be estimated via a least-squares minimization relative
to the empirical waiting-time distribution for the ambulance-vehicle queue
as determined by Monte Carlo (MC) simulation.

\subsection{Statistical Tests}
\label{WaitTest}
We compare the exact and approximate ambulance-vehicle waiting-time distributions with simulation results
in two ways:
(i) We apply a null-hypothesis test in the spirit of a
one-sample Kolmogorov-Smirnov (KS) test to each of the theoretical distributions.
(ii) We test the two theoretical distributions against each other via a likelihood-ratio (LR) test
to ascertain which has a better goodness-of-fit (GOF).

\begin{figure}
\FIGURE
%\begin{center}
{\includegraphics[width=\wsclx\linewidth, height=\hsclx\linewidth]{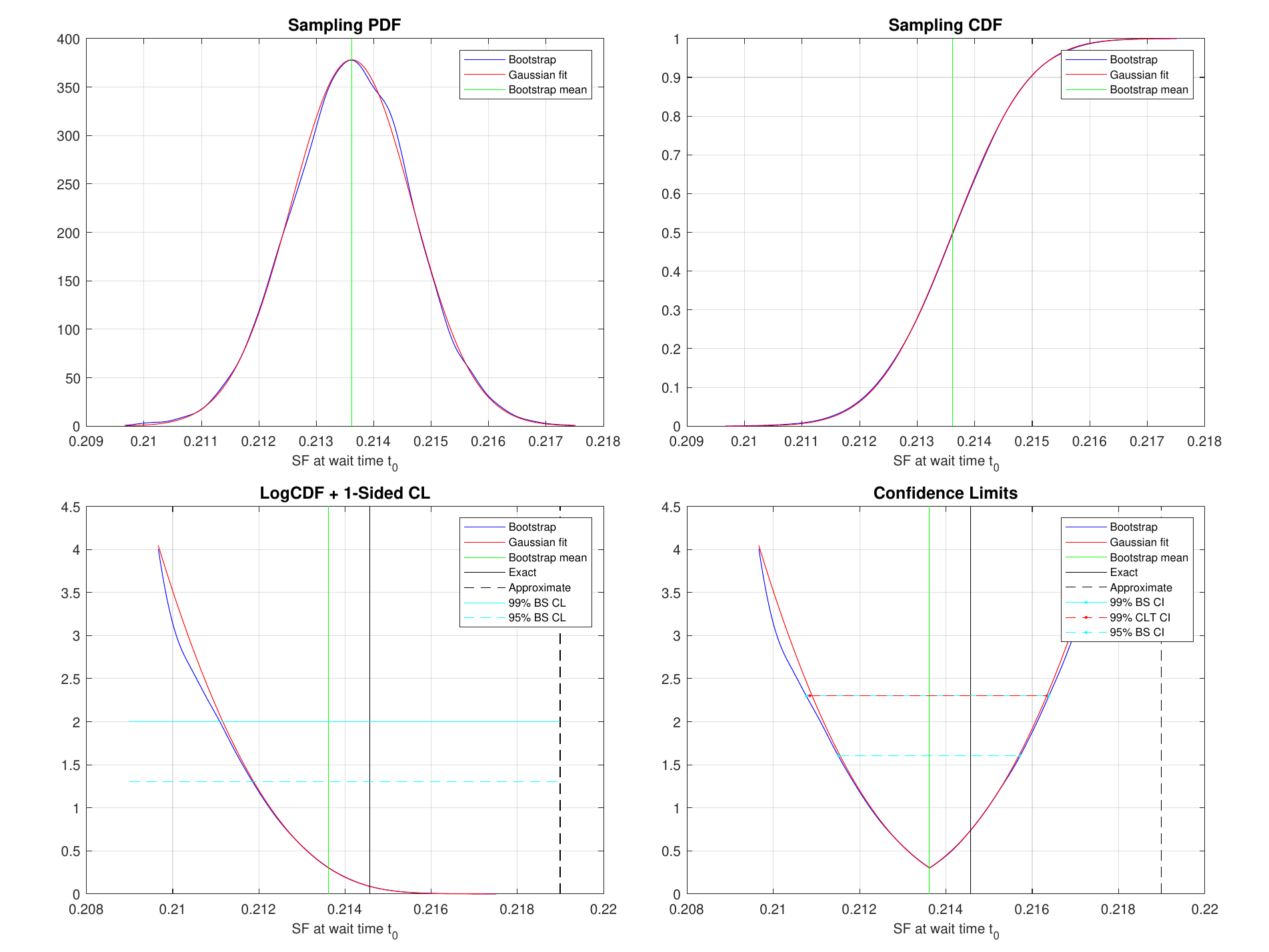}}
{\hphantom{x}\label{fig:AmboWaitTimeCI}}
{Confidence interval for the ambulance-vehicle waiting-time null-hypothesis test,
     generated via the bootstrap technique}
%\end{center}
\end{figure}

In the null-hypothesis test, we consider the waiting time
\mbox{$t = t_0$}
at which the discrepancy between the theoretical exact and approximate SFs
\mbox{$|\bar{F}^{\text{c}}_{\text{ex}}(t) - \bar{F}^{\text{c}}_{\text{apx}}(t)|$}
is maximum.
At this point, the PDFs are equal, so that
\mbox{$P^{\text{c}}_{\text{ex}}(t_0) - P^{\text{c}}_{\text{apx}}(t_0) = 0$}
yields the equation to be solved for $t_0$.
We then consider the simulation-derived empirical SF
\mbox{$\bar{F}^{\text{c}}_{\text{emp}}(t)$}
at
\mbox{$t = t_0$},
and ask whether the null hypotheses that
\mbox{$\bar{F}^{\text{c}}_{\text{ex}}(t_0) = \bar{F}^{\text{c}}_{\text{emp}}(t_0)$}
and
\mbox{$\bar{F}^{\text{c}}_{\text{apx}}(t_0) = \bar{F}^{\text{c}}_{\text{emp}}(t_0)$}
can be rejected at some confidence level $1-\alpha$.
For a test of sufficiently high power, one expects to be able to reject the latter null hypothesis,
but not concurrently the former.
The steady-state MC simulation of the ambulance-vehicle waiting time is considered within the
framework of regenerative simulation \citep{NP:Crane74}, and bootstrapping \citep{NP:Hesterberg11}
is used to construct the confidence interval (CI)
\mbox{$[y_{\text{lo}},y_{\text{hi}}]$}
for
\mbox{$\bar{F}^{\text{c}}_{\text{emp}}(t_0)$}.
The regeneration points are taken to be the epochs at which the system becomes empty.
The number of bootstrap resamplings generated is
\mbox{$B = 10^4$},
and we set
\mbox{$\alpha = 0.01$}.
The power of the test is governed by the stop time of the steady-state simulation.
We compute the Booleans
\begin{equation}
h_{\text{ex}}  \equiv I(\bar{F}^{\text{c}}_{\text{ex}}(t_0) \notin [y_{\text{lo}},y_{\text{hi}}]) \,, \quad
h_{\text{apx}} \equiv I(\bar{F}^{\text{c}}_{\text{apx}}(t_0) \notin [y_{\text{lo}},y_{\text{hi}}]) \,,
\end{equation}
where $I(\mathcal{A})$ denotes the indicator function such that
\mbox{$I(\mathcal{A}) = 1$}
if the proposition $\mathcal{A}$ is true, and is zero otherwise.
If we repeat the computation over many MC simulation runs, the quantities
\mbox{$h_{\text{ex}}, h_{\text{apx}}$}
become Bernoulli random variables (RVs).
Since we know {\it a priori} that
\mbox{$\bar{F}^{\text{c}}_{\text{ex}}(t)$}
is exact while
\mbox{$\bar{F}^{\text{c}}_{\text{apx}}(t)$}
is not, an estimate of the false-alarm rate (FAR) is given by the mean $\bar{h}_{\text{ex}}$
and we expect that
\mbox{$\bar{h}_{\text{ex}} \simeq \alpha$}.
This constitutes a sanity check on the simulations.
On the other hand, the missed detection rate (MDR) is estimated by
\mbox{$1 - \bar{h}_{\text{apx}}$},
which means that the value of
\mbox{$\bar{h}_{\text{apx}}$}
can be used to gauge the power of the test.
Figure~\ref{fig:AmboWaitTimeCI} displays the results for a steady-state simulation with a
stop time of $10^6$, yielding $637$ regeneration cycles.
The first three graphs display the PDF, cumulative distribution function (CDF) and (base-10) log-CDF of the bootstrap
generated sampling distribution, along with a Gaussian fit. In order to conveniently
display the two-sided confidence limits (CL) on the log-CDF graph, we plot
an everywhere positive version of the
function obtained by reflecting about the horizontal at the mean level.
This is presented in the last graph, where CIs derived from
both bootstrapping and the central limit theorem \citep{NP:Crane74} are indicated.
The solid black line gives the exact result at the test point $t_0$, and the dashed
black line is the approximation at the test point.

A regeneration cycle in the steady-state simulation comprises the union of an idle period
with an adjacent partial busy period \citep{NP:Artalejo01}.
A further sanity check of the simulation is obtained by constructing the empirical
busy-period distribution and confirming its agreement with the theoretical results
derived in \cite{NP:Zuk23B}.

\begin{table}
\TABLE
%\centering
{Null-Hypothesis Test\label{tab:nullhyp}}
{\begin{tabular}{|c|c|c|c|}
\hline
$\nu_{\text{amb}}$ & Stop Time & FAR & MDR \\
\hline\hline
$1$   & $10^5$         & $0.010$ & $0.900$ \\
$1$   & $3\times 10^5$ & $0.005$ & $0.600$ \\
$1$   & $10^6$         & $0.010$ & $0.040$ \\
$2/3$ & $10^6$         & $0.014$ & $0.960$ \\
\hline
\end{tabular}}
{}
\end{table}

% Created with AmboWaitTimeComboTest.m
% Random number seed: sd = 1234523

For the LR test, we consider the log likelihood ratio (LLR)
\begin{equation}
\Lambda \equiv \frac{1}{N_{\text{sim}}}\sum_{k= 1}^{N_{\text{sim}}}
     \ln\left[\frac{P^{\text{c}}_{\text{ex}}(t_k)}{P^{\text{c}}_{\text{apx}}(t_k)}\right] \,,
\end{equation}
where the $t_k$,
\mbox{$k = 1,2,\ldots,N_{\text{sim}}$},
are the $N_{\text{sim}}$ (non-zero) ambulance-vehicle waiting times generated by the
steady-state MC simulation.
As with the previous test, the steady-state MC simulation is
considered within the framework of regenerative simulation and the bootstrapping technique, with
\mbox{$B = 10^4$}
bootstrap resamplings,
is used to construct the CI
\mbox{$[\lambda_{\text{lo}},\lambda_{\text{hi}}]$}
for $\Lambda$.

The following observations can be made about the LR test:
(1) The `exact' PDF is deemed to be a statistically significantly
better fit to the simulation data than the `approximate' PDF
if the CI of the LLR lies entirely to the right of zero,
{\it i.e.}\ the LLR is sufficiently positive.
(2) If the CI lies entirely to the left of zero, this is
a false alarm, as it would indicate the `approximate' PDF is
statistically significant as a better fit.
(3) If zero lies within the CI, then we have a missed
detection, since the data is consistent with both PDFs
being equally acceptable fits.
Thus, we compute the Booleans
\begin{equation}
h_{\text{fa}} \equiv \mathcal{I}(\lambda_{\text{hi}} < 0) \,, \;
h_{\text{md}} \equiv \mathcal{I}(\lambda_{\text{lo}} < 0 \BoolAnd \lambda_{\text{hi}} \geq 0) \;,
\end{equation}
that track false alarms and missed detections, respectively.
If we repeat the computation over many MC simulation runs $n_{\text{runs}}$, the quantities
\mbox{$h_{\text{fa}}, h_{\text{md}}$}
become Bernoulli RVs, with the FAR and MDR given by their empirical means
\mbox{$\bar{h}_{\text{fa}}, \bar{h}_{\text{md}}$},
respectively.

\begin{figure}
\FIGURE
%\begin{center}
{\includegraphics[width=\wsclx\linewidth, height=\hsclx\linewidth]{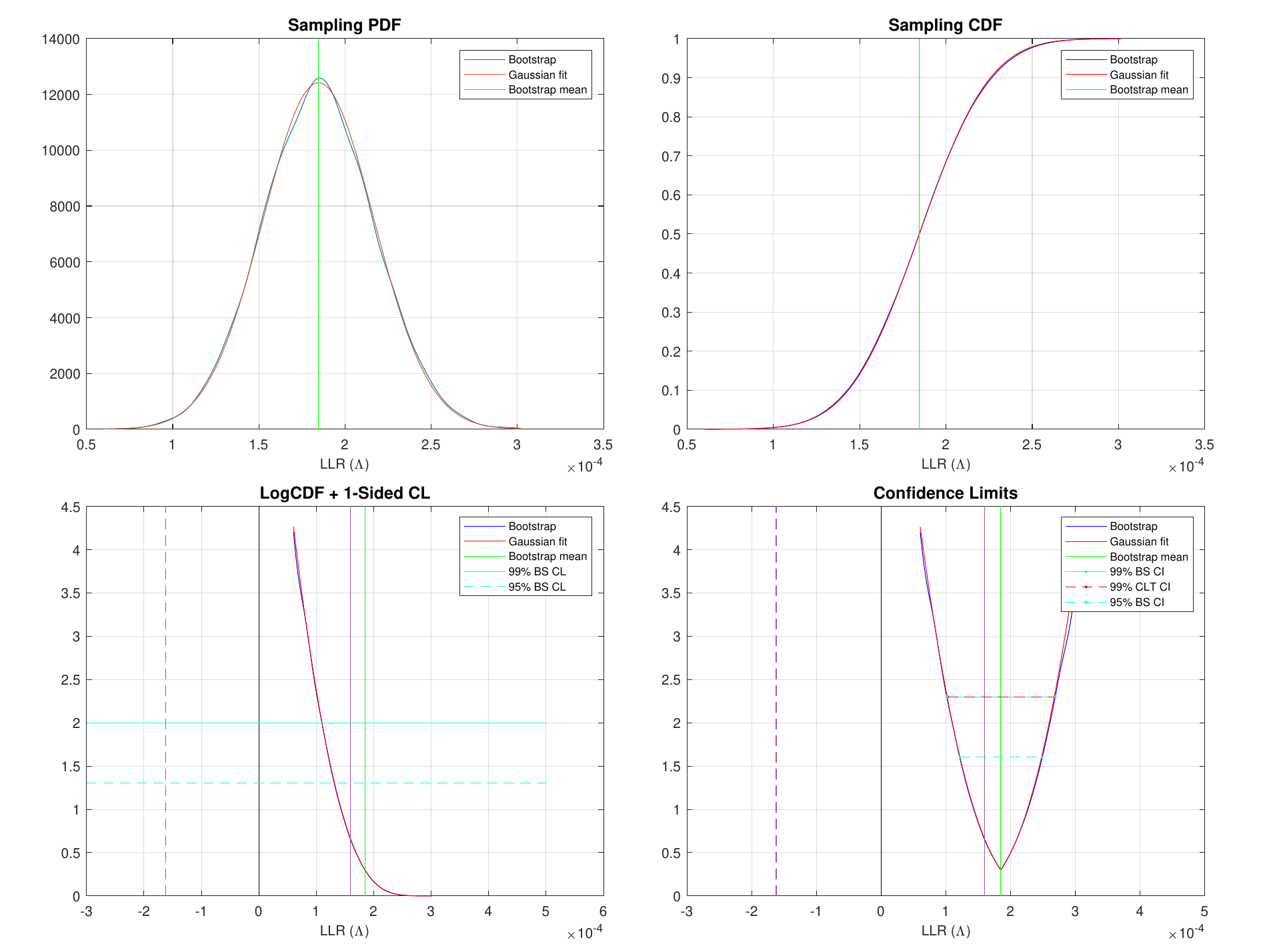}}
{\hphantom{x}\label{fig:AmboWaitTimeLRCI}}
{Confidence interval for the ambulance-vehicle waiting-time likelihood-ratio test,
     generated via the bootstrap technique.}
%\end{center}
\end{figure}

Figure~\ref{fig:AmboWaitTimeLRCI} displays the results for a steady-state simulation with a
stop time of $10^6$, yielding $637$ regeneration cycles.
The schema for the graphs follows that used with the previous null-hypothesis test.
We have also computed and plotted the positions of the Kullback-Leibler (KL) divergences
\begin{equation}
d_1 \equiv \int_0^\infty dt\, P^{\text{c}}_{\text{ex}}(t)
     \ln\left[\frac{P^{\text{c}}_{\text{ex}}(t)}{P^{\text{c}}_{\text{apx}}(t)}\right] \,, \quad
d_2 \equiv \int_0^\infty dt\, P^{\text{c}}_{\text{apx}}(t)
     \ln\left[\frac{P^{\text{c}}_{\text{apx}}(t)}{P^{\text{c}}_{\text{ex}}(t)}\right] \,.
\end{equation}
The solid lilac line represents $d_1$, while the dashed lilac line represents the negative of $d_2$.
One may note that the LR is an empirical estimator for $d_1$ under the assumption that
\mbox{$P^{\text{c}}_{\text{ex}}(t)$}
is indeed an exact result.
Likewise, the LR is an empirical estimator for $-d_2$ under the assumptions that
\mbox{$P^{\text{c}}_{\text{apx}}(t)$}
is actually an exact result.
We are able to reject this latter hypothesis if $-d_2$ lies outside the CI.
Otherwise, we have a missed detection.
Similarly, $d_1$ lying outside the CI constitutes a false alarm.
Consequently, we may construct the alternative Booleans
\begin{equation}
k_\text{fa} \equiv \mathcal{I}(d_1 \notin [\lambda_{\text{lo}},\lambda_{\text{hi}}]) \,, \;
k_\text{md} \equiv \mathcal{I}(-d_2 \in [\lambda_{\text{lo}},\lambda_{\text{hi}}]) \,,
\end{equation}
that track false alarms and missed detections, respectively.

The results over multiple MC runs,
\mbox{$n_{\text{runs}} = 200$},
and various model and simulation parameter values are presented in Table~\ref{tab:lrtest}.
Vanishing statistics indicate values of less than $1/n_{\text{runs}}$.
We see that, for
\mbox{$\nu_{\text{amb}} = 1$},
a steady-state simulation stop time of $10^6$ is required to differentiate between the exact and approximate
distributions, whereas, for
\mbox{$\nu_{\text{amb}} = 2/3$}
with the same stop time, one is unable to conclude that the approximate distribution
performs worse than the exact one.
The final two columns are the results for the KL version of the test. We see that it has
greater power, but is considerably more complex to compute.

\begin{table}
\TABLE
%\centering
{Likelihood-Ratio Test\label{tab:lrtest}}
{\begin{tabular}{|c|c||c|c||c|c|}
\hline
$\nu_{\text{amb}}$ & Stop Time & FAR & MDR & FAR-KL & MDR-KL \\
\hline\hline
$1$   & $10^5$         & $0.000$ & $0.758$ & $0.025$ & $0.283$ \\
$1$   & $3\times 10^5$ & $0.000$ & $0.375$ & $0.009$ & $0.000$ \\
$1$   & $10^6$         & $0.000$ & $0.005$ & $0.000$ & $0.000$ \\
$2/3$ & $10^6$         & $0.000$ & $0.930$ & $0.020$ & $0.815$ \\
\hline
\end{tabular}}
{}
\end{table}

% Created with AmboWaitTimeComboTest.m
% Random number seed: sd = 1234523

The upshot of the foregoing analysis is that one must observe
\mbox{$N\mu T_{\text{stop}} = 10^7$}
patients being processed in the system, for the worst-case parameters, in order to be
able to discriminate between the exact and approximate distributions, and
even this number in not sufficient for other parametric cases.

\section{Results}
\label{Results}
We introduce the SF
\mbox{$\bar{F}(n)$}
associated with the PMF
\mbox{$P(n)$}
as
\begin{equation}
\bar{F}(n) \equiv \Pr(L > n) = 1 - \sum_{k=0}^{n} P(k) \,.
\label{SFDiscrete}
\end{equation}
Then the mean is given by
\begin{equation}
\bar{L} \equiv \sum_{n=0}^\infty nP(n) = \sum_{n=0}^\infty \bar{F}(n) \,.
\end{equation}
To account for the discontinuity at
\mbox{$n = 0$},
an exponential ({\it i.e.}\ geometric) fit to
\mbox{$\bar{F}(n)$}
has the form
\begin{equation}
\bar{F}_{\text{fit}}(n) = \left\{
\begin{array}{ccl}
1                & \quad\text{for}\quad & n < 0 \,, \\
\bar{F}(0)\rho^n & \quad\text{for}\quad & n = 0,1,2,\ldots \,,
\end{array}
\right.
\label{SFGeom}
\end{equation}
noting that
\mbox{$\bar{F}(0) = 1 - P(0)$}.
Fitting the mean yields
\mbox{$\rho = 1 - \bar{F}(0)/\bar{L}$}.
Therefore,
\mbox{$H(n) \equiv -\log_{10}\bar{F}_{\text{fit}}(n)$}
is perfectly linear in
\mbox{$n \ge 0$}
for geometric PMFs with a discontinuity at the origin.

The offload delay rate RV is represented by the compound RV
\begin{equation}
\Omega = (30/T_{\text{mth}}){\cdot}\sum_{n = 1}^{N_{\text{amb}}} W_{\text{amb}}^{(n)}
     = 30\lambda_{\text{amb}}{\cdot}\frac{1}{\overline{N}_{\text{amb}}}
     \sum_{n = 1}^{N_{\text{amb}}} W_{\text{amb}}^{(n)} \;,
\end{equation}
where $N_{\text{amb}}$ is the Poisson RV for the number of ambulance arrivals in one month,
$T_{\text{mth}}$ denotes a time interval of one month and, by definition, we have
\mbox{$\overline{N}_{\text{amb}}/T_{\text{mth}} = \lambda_{\text{amb}}$}.
It follows that the average offload delay rate is given by
\mbox{$\overline{\Omega} = 30\lambda_{\text{amb}}\overline{W}_{\text{amb}} = 30\overline{L}_{\text{amb}}$}.
Since
\mbox{$N_{\text{amb}} \gg 1$}
almost surely, the central limit theorem (CLT) implies that
\mbox{$\Omega \simeq (30\overline{L}_{\text{amb}}/\overline{N}_{\text{amb}}){\cdot}N_{\text{amb}}$}.
In characterizing the offload delay rate, our idealized modelling does not account for the expected
`undelayed' patient offload time. We are, in effect, assuming that it is a relatively small fraction
of the mean treatment time.
Alternatively, if one were modelling the wider emergency services network,
this could be included by adding a constant offset to the ambulance transit time.

An exponential ansatz for the average offload delay rate,
whose derivation is provided in Appendix~\ref{Ansatz}, is given by
\begin{equation}
\overline{L}_{\text{amb}} \simeq \overline{L}_{\text{hi}} + \overline{L}\vphantom{L}_{\text{med}}^{\text{amb}}{\cdot}t^M \,,
     \quad t \equiv 1 - (1 - P_{\text{med}}^{\text{amb}})/\overline{L}\vphantom{L}_{\text{med}}^{\text{amb}} \,, \quad
     \overline{L}\vphantom{L}_{\text{med}}^{\text{amb}} = (1-q)\overline{L}_{\text{med}} \;,
\label{LAmbAnsatz}
\end{equation}
with
\begin{equation}
P_{\text{med}}^{\text{amb}} \equiv P_{\text{med}}(0) + [1 - P_{\text{med}}(0)]
     {\cdot}\frac{q(1-t')}{1 - qt'} \,,
     \quad t' \equiv 1 - [1 - P_{\text{med}}(0)]/\overline{L}_{\text{med}} \,,
\end{equation}
where $q$ denotes the fraction of intermediate-priority arrivals that are walk-ins,
as given by (\ref{pq}).
The mean queue lengths
\mbox{$\overline{L}_{\text{med}}, \overline{L}_{\text{hi}}$}
refer to those for all intermediate-priority and high-priority ambulance arrivals, respectively,
as given below (\ref{LM}),
having been rendered unconditional according to
\begin{equation}
\overline{L}_{\text{med}} = (1-P_{\text{NW}})\overline{L}\vphantom{L}^{\text{c}}_{\text{med}} \;, \quad
\overline{L}_{\text{hi}}  = (1-P_{\text{NW}})\overline{L}\vphantom{L}^{\text{c}}_{\text{hi}}\;.
\end{equation}
Here, the quantity
\mbox{$P_{\text{med}}(0)$}
refers to the PMF at zero queue length for \emph{all} intermediate priority arrivals,
and is estimated as
\mbox{$P_{\text{med}}(0) = P_2(0)$},
setting
\mbox{$\kappa = 2$}
in
\begin{equation}
P_\kappa(0) \simeq 1 - \frac{2(\overline{L}\vphantom{L}_\kappa)^2}
     {\overline{L}\vphantom{L}_\kappa + \overline{(L_\kappa)^2}} \;,
\end{equation}
where the second moment follows from (\ref{SecondMom}) and
\begin{equation}
\overline{L}_\kappa      = (1-P_{\text{NW}})\overline{L}\vphantom{L}^{\text{c}}_\kappa \;, \quad
\overline{(L_\kappa)^2} = (1-P_{\text{NW}})\overline{(L^{\text{c}}_\kappa)^2} \;.
\end{equation}

\begin{figure}
\FIGURE
%\begin{center}
{\includegraphics[width=\wscl\linewidth, height=\hscl\linewidth]{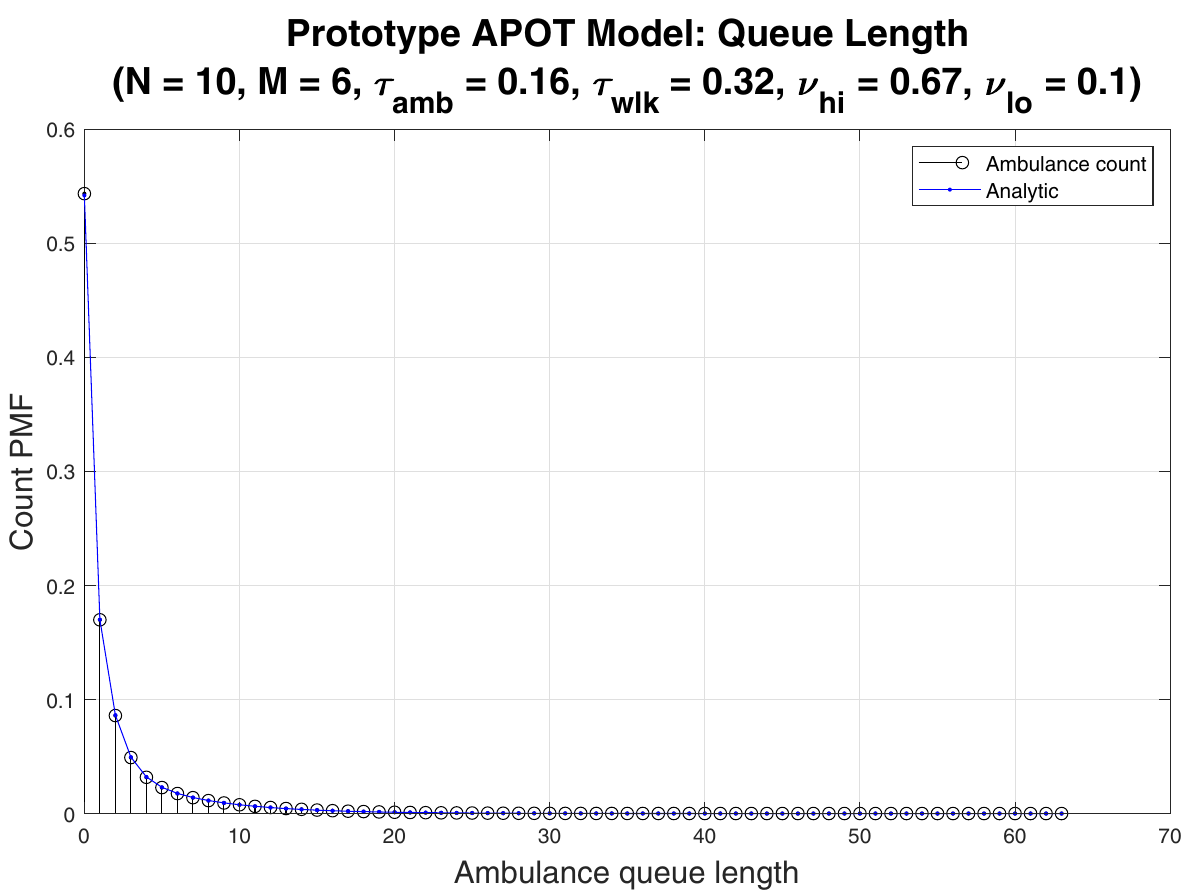}}
{\hphantom{x}\label{fig:QueueSizePMF}}
{Ambulance-vehicle queue-length PMF with $r = 0.95$, $\nu_{\text{amb}} = 2/3$
     and simulation stop time $T_{\text{stop}} = 10^6$.}
%\end{center}
\end{figure}

\begin{figure}
\FIGURE
%\begin{center}
{\includegraphics[width=\wscl\linewidth, height=\hscl\linewidth]{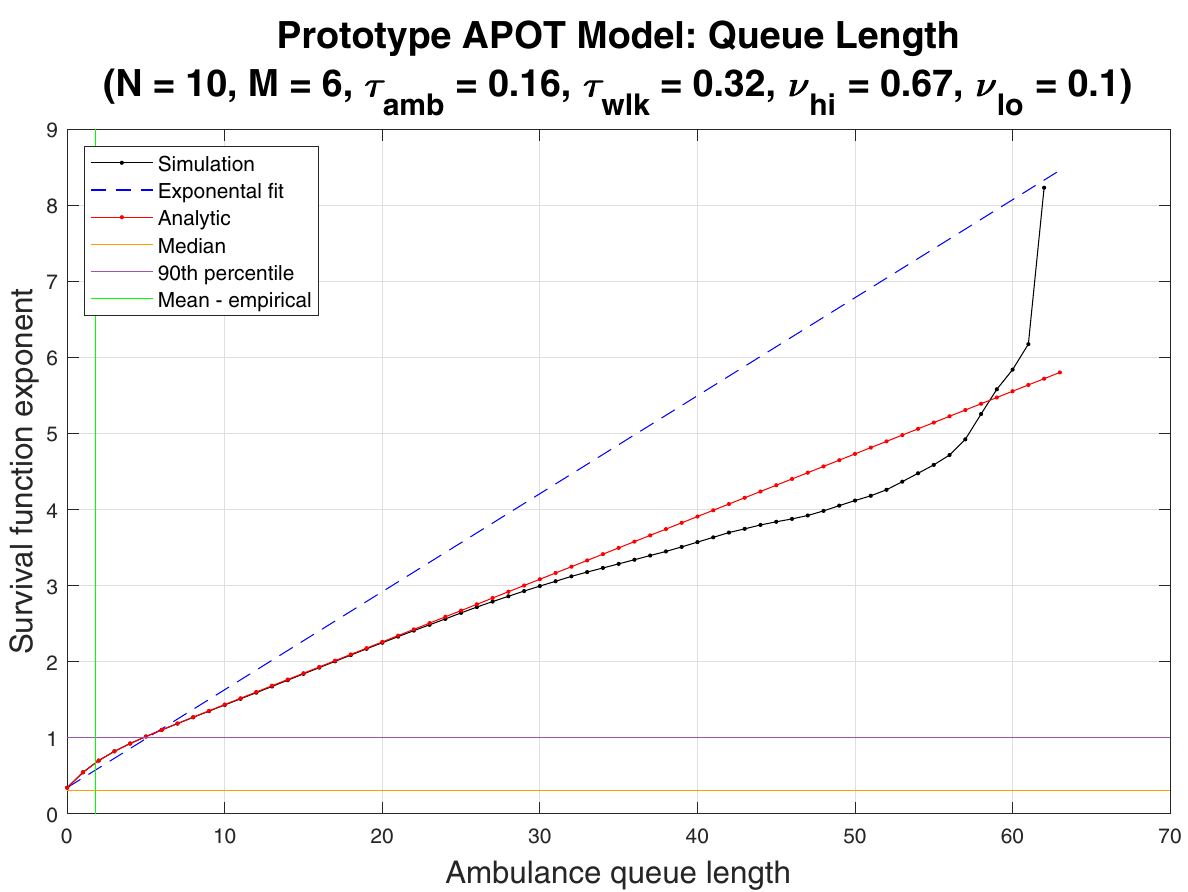}}
{\hphantom{x}\label{fig:QueueSizeSF}}
{Log-survival function for the ambulance-vehicle queue-length with $r = 0.95$, $\nu_{\text{amb}} = 2/3$
     and simulation stop time $T_{\text{stop}} = 10^6$.}
%\end{center}
\end{figure}

\begin{figure}
\FIGURE
%\begin{center}
{\includegraphics[width=\wscl\linewidth, height=\hscl\linewidth]{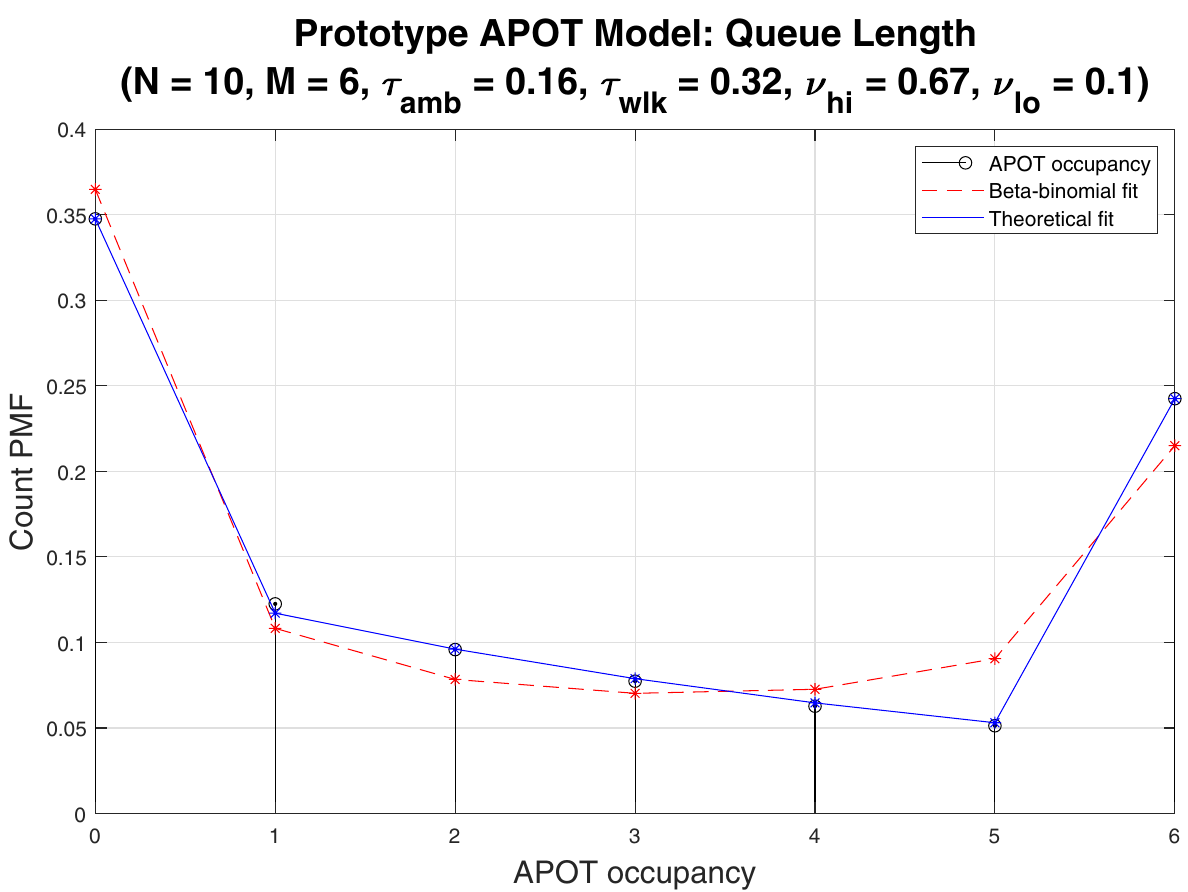}}
{\hphantom{x}\label{fig:QueueSizeAPOT}}
{PMF for the APOT occupancy with $r = 0.95$, $\nu_{\text{amb}} = 2/3$
     and simulation stop time $T_{\text{stop}} = 10^6$.}
%\end{center}
\end{figure}

\begin{figure}
\FIGURE
%\begin{center}
{\includegraphics[width=\wscl\linewidth, height=\hscl\linewidth]{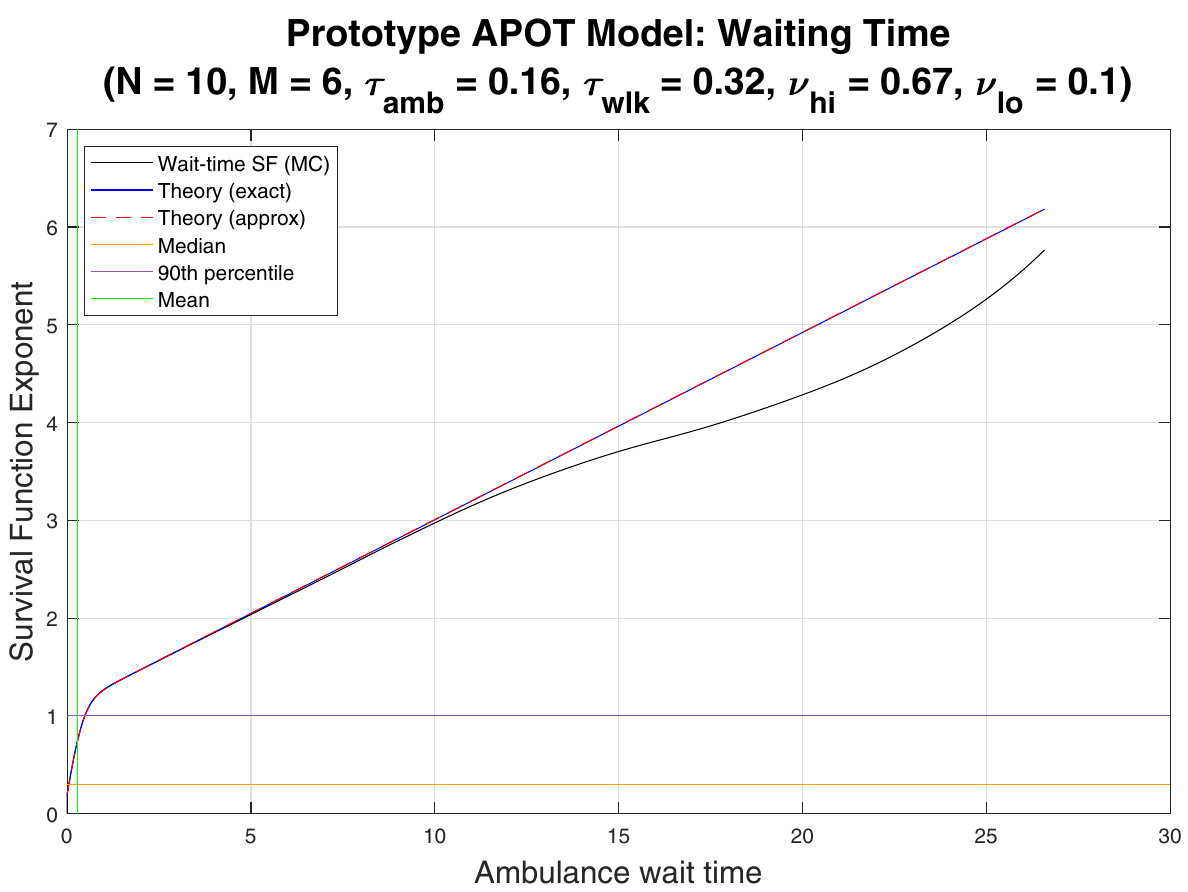}}
{\hphantom{x}\label{fig:WaitTime}}
{Log-survival function for the ambulance-vehicle waiting time, with $r = 0.95$, $\nu_{\text{amb}} = 2/3$
     and simulation stop time $T_{\text{stop}} = 10^6$.}
%\end{center}
\end{figure}

\begin{figure}
\FIGURE
%\begin{center}
{\includegraphics[width=\wscl\linewidth, height=\hscl\linewidth]{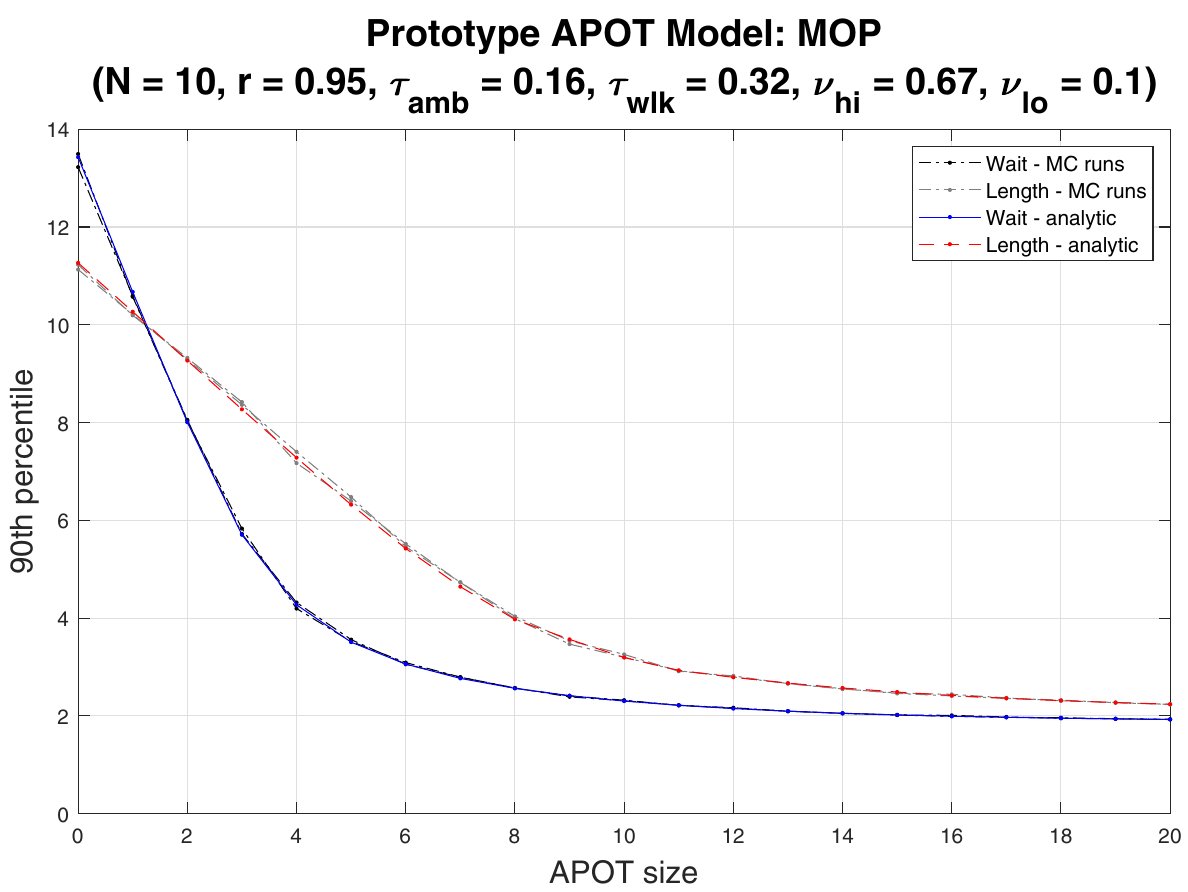}}
{\hphantom{x}\label{fig:90PC}}
{90th percentiles versus APOT size, with $\nu_{\text{amb}} = 2/3$ and simulation stop time
     $T_{\text{stop}} = 10^6$.}
%\end{center}
\end{figure}

\begin{figure}
\FIGURE
%\begin{center}
{\includegraphics[width=\wscl\linewidth, height=\hscl\linewidth]{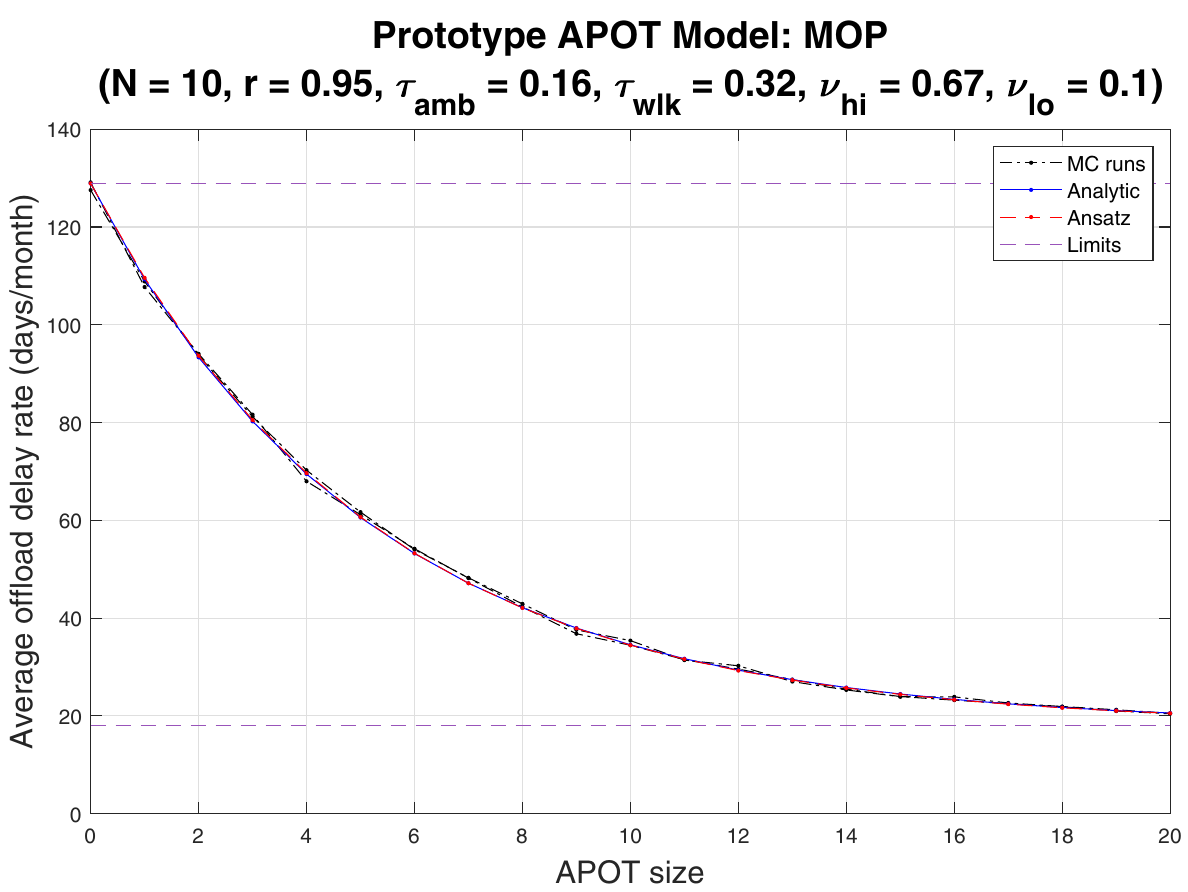}}
{\hphantom{x}\label{fig:Rate}}
{Average offload delay rate versus APOT size, with $\nu_{\text{amb}} = 2/3$ and simulation stop time
     $T_{\text{stop}} = 10^6$.}
%\end{center}
\end{figure}

In Figure~\ref{fig:QueueSizePMF}, the black stem plot presents the results of the MC simulation
for the ambulance-vehicle queue-length PMF, which is seen to be in very good agreement with
the blue curve generated by the exact theoretical distribution.
The simulation stop time was set to
\mbox{$T_{\text{stop}} = 10^6$}
(measured in units of the mean treatment time).
We also set
\mbox{$r = 0.95, \nu_{\text{amb}} = 2/3$}
in accordance with Table~\ref{tab:parm},
with other parameter values as indicated on the figure.
The same choices apply to Figures~\ref{fig:QueueSizeSF}--\ref{fig:Rate}.
Figure~\ref{fig:QueueSizeSF} replots the same distribution as the negative base-10 logarithm
of the SF computed according to (\ref{SFDiscrete}).
The black line is the empirical result from simulation, while
the red curve is the exact theoretical result.
For reference,
the dashed blue line represents the fit to an exponential ({\it i.e.}\ geometric) distribution
according to (\ref{SFGeom}).
In Figure~\ref{fig:QueueSizeAPOT}, the black stem plot presents the results of the MC simulation
for the APOT occupancy. The blue curve is the exact theoretical result, which is in excellent agreement
with the simulation.
The dashed red curve represents a fit of the empirical ({\it i.e.}\ simulation)
data to a beta-binomial distribution.
For many parameter sets, this results in very good agreement; considerably better than that
obtained for the current example.

In Figure~\ref{fig:WaitTime}, we plot the log-SF
\mbox{$H_{\text{W}}(t) \equiv -\log_{10}\bar{F}_{\text{W}}(t)$}
for the ambulance-vehicle waiting time.
The black curve is the empirical result from MC simulation.
The exact (blue) and approximate (red) theoretical curves agree within the linewidth of the graph
and are consistent with the simulation results.

% Created with PlotQueueSize.m
% Random number seed: sd = xxxxx
% Tstop = 1e6, r = 0.95, AmbFrac = 2/3
% RunTime = 16 min

% Created with PlotQueueSize.m
% Random number seed: sd = xxxxx
% Tstop = 1e6, r = 0.95, AmbFrac = 2/3
% RunTime = 16 min

% Created with PlotQueueSize.m
% Random number seed: sd = xxxxx

% Created with PlotAmboWaitTime.m
% Random number seed: sd = xxxxx
% Tstop = 1e6, r = 0.95, AmbFrac = 2/3
% RunTime = 16 min

In Figure~\ref{fig:90PC}, we plot the exact theoretical results for
90th percentiles of the ambulance-vehicle
queue length (red curve) and scaled waiting time (blue curve) as functions of the APOT size $M$.
By multiplying the waiting time by the mean arrival rate
$\lambda_{\text{amb}}$, it becomes dimensionless and is of the same scale as the queue
length.
The overlaid simulation results from two independent runs,
comprising the dashed black curve for the
waiting time and the dashed grey curve for the queue length, are totally consistent
with the theoretical expectations.
In Figure~\ref{fig:Rate}, we plot the average offload delay rate as a function of the
APOT size $M$. The exact analytic curve (blue) agrees with the exponential ansatz (red)
to within the linewidth of the graph, and they are consistent with results from MC
simulation (black) displayed for two independent runs.
The mean ambulance-vehicle queue length versus APOT size is obtained from Figure~\ref{fig:Rate}
by dividing the vertical axis by 30, while the mean waiting time
in units of the mean treatment time is obtained by dividing by
\mbox{$30\lambda_{\text{amb}} = 30/\tau_{\text{amb}} = 190$}.
The limiting value as the APOT size $M$ tends to infinity
reflects the fact that high-priority patients are assumed not to be offloaded into the APOT.
It is important to note that the large numbers associated with these figures are not intended
to be indicative of any actual real-world system, but result from model parameters chosen to
stress the numerical algorithms, and thus confirm their robustness.

% Created with PlotMOPVsAPOTSize.m
% Random number seed: sd = xxxxx

% Created with PlotMOPVsAPOTSize.m
% Random number seed: sd = xxxxx

\section{Conclusions}
\label{Concl}
The effect of APOT/OZ size on ambulance ramping has been studied within a simple prototype model
that aims to capture the most important characteristics of the ambulance/ED interface
while being amenable to analytical methods.
We have obtained a closed-form, purely algebraic expression that approximates
the dependence of the average offload delay rate
({\it i.e.}\ the number of ambulance days lost per month due to offload delay)
on the APOT size,
and found that it is, for all practical purposes,
indistinguishable from the exact result,
which has also been derived.
Since the gradient decreases monotonically to zero,
it is observed that there is no `sweet spot' in this relationship:
While the aggregated benefit can be substantial,
the incremental benefit associated with each additional APOT place decreases
with each such place.

Future work could be directed at generalizing the queue discipline.
In the current model, patients are admitted to the ED according only to their priority level,
independent of their ED arrival source (i.e., waiting room for walk-ins, ambulance-vehicle queue
or OZ). However, it has been recorded in some trials that ED staff have prioritized the
waiting room over the OZ, leading to sub-optimal outcomes \citep{NP:Carter15,NP:Laan16}.
It has also been suggested that
that ED beds should be given to OZ patients at least 35\% of the time in order for for the OZ
to have a positive impact on AOD \citep{NP:Laan16}.
The model could be extended to capture and compare these effects.

Another generalization would be to remove the simplifying assumptions that all high-priority
patients arrive by ambulance and all low-priority patents are walk-ins.
Additionally, the number of priority levels could be increased.
This would make it easier to fit model parameters to real-world data.
Analytical results would still be possible by appealing to the explicit results for
the general multi-level priority problem considered in \citep{NP:Zuk23A}.

Finally, one could model an APOT that serves more than a single acuity level and
is not of constant size in patient places, but is constant in staffing level.
In this case, the number of patients that can be accommodated at any given time
depends on the mixture of acuity levels, as determined by required staff-to-patient ratios.

\begin{APPENDICES}
\section{Marginal PGFs}
\label{MargPGF}
We consider a $K$-level non-preemptive Markovian priority queue with priority levels
labelled
\mbox{$\kappa = 1,2,\ldots,K$}
in increasing order of priority, so that
\mbox{$\kappa = 1$}
represents the highest priority level.
We let $r_\kappa$ denote the level traffic intensity for priority level $\kappa$,
and introduce the cumulative sums
\begin{equation}
\sigma_\kappa \equiv \sum_{j=1}^\kappa r_j \;.
\end{equation}
Let
\mbox{$g^{(\kappa)}(z)$}
denote the PGF for the wait-conditional marginal queue length distribution
corresponding to priority level $\kappa$,
so that
\mbox{$g^{(\kappa)}(z) = \sum_{n=1}^\infty P^{\text{c}}_\kappa(n)z^n$}.
Following \citep{NP:Zuk23C,NP:Zuk23,NP:Zuk23A},
based on the approach of \citet{NP:Cohen56},
we have, for the highest priority level,
\begin{equation}
g^{(1)}(z) = \frac{1 - r_1}{1 - r_1 z} \;.
\label{g-one}
\end{equation}
For all other priority levels,
\begin{equation}
g^{(\kappa)}(z) = \frac{1 - \sigma_\kappa}{\zeta^{(\kappa)}_+(z) - \sigma_\kappa} \;,
\label{g-kappa}
\end{equation}
for
\mbox{$\kappa = 2,3,\ldots,K$},
where
\begin{equation}
\zeta^{(\kappa)}_\pm(z) = \half\left[b_\kappa(z) \pm
     \xi_\kappa(z)\sqrt{\Delta_\kappa(z)}\right] \;,
\end{equation}
with
\begin{equation}
b_\kappa(z) \equiv 1 + \sigma_\kappa - r_\kappa z \,, \;
     \Delta_\kappa(z) \equiv b^2_\kappa(z) - 4\sigma_{\kappa-1} \;.
\end{equation}
The square root is taken to be the principal-value branch, and the sign function
\mbox{$\xi_\kappa(z)$},
which serves to make
\mbox{$\zeta^{(\kappa)}_\pm(z)$}
analytic in the entire complex $z$-plane away from the real axis, is given by
\begin{equation}
\xi_\kappa(z) = \left\{
\begin{array}{ccc}
+1 & \quad\text{for}\quad & \Rez z \leq (x^{(\kappa)}_+ + x^{(\kappa)}_-)/2 \\
-1 & \quad\text{for}\quad & \Rez z > (x^{(\kappa)}_+ + x^{(\kappa)}_-)/2
\end{array}
\right. \;,
\label{xiz}
\end{equation}
where
\begin{equation}
x^{(\kappa)}_\pm \equiv 1 + \left(1 \pm \sqrt{\sigma_{\kappa-1}}\right)^2/r_\kappa \;.
\end{equation}
These positive real values represent the upper and lower limits of the cut in the $z$-plane
for the $\kappa$-th priority level problem.
One may note that (\ref{g-one}) and (\ref{g-kappa}) are consistent under the identification
\mbox{$\sigma_0 \equiv 0$}.

\section{Means and Moments}
\label{Means}
If
\mbox{$G(z)$}
denotes the PGF for a marginal queue-length distribution,
then the mean queue length is given by
\begin{equation}
\overline{L} = \left.\frac{d}{dz}G(z)\right|_{z=1} \;.
\end{equation}
We recall that, in the present application, the PGF for the wait-conditional
queue-length marginal for priority level $\kappa$ is given by (\ref{g-kappa}).
The requirement that
\mbox{$g^{(\kappa)}(1) = 1$}
implies that
\mbox{$\zeta^{(\kappa)}_+(1) = 1$}.
Thus, we have for the associated mean queue length
\begin{equation}
\overline{L}\vphantom{L}^{\text{c}}_\kappa \equiv  \left.\frac{d}{dz}g^{(\kappa)}(z)\right|_{z=1}
     = -\frac{\zeta^{(\kappa)\prime}_+(1)}{1-\sigma_\kappa} \;.
\end{equation}
The quadratic equation satisfied by
\mbox{$\zeta^{(\kappa)}_\pm(z)$}
implies, for the derivative with respect to $z$, that
\begin{equation}
\zeta^{(\kappa)\prime}_+(1) = \frac{r_\kappa\zeta^{(\kappa)}_+(1)}
     {1 + \sigma_\kappa - r_\kappa - 2\zeta^{(\kappa)}_+(1)}
     = -\frac{r_\kappa}{1 - \sigma_{\kappa-1}} \;.
\end{equation}
Consequently,
\begin{equation}
\overline{L}\vphantom{L}^{\text{c}}_\kappa = \frac{r_\kappa}{(1-\sigma_\kappa)(1-\sigma_{\kappa-1})} \;.
\end{equation}
With the aid of Little's law
\mbox{$\overline{L}\vphantom{L}^{\text{c}}_\kappa = \lambda_\kappa\overline{W}\vphantom{W}^{\text{c}}_\kappa$},
we obtain for the mean waiting time
\begin{equation}
\overline{W}\vphantom{W}^{\text{c}}_\kappa = \frac{1}{N\mu}{\cdot}
     \frac{1}{(1-\sigma_\kappa)(1-\sigma_{\kappa-1})} \;,
\end{equation}
noting that
\mbox{$r_\kappa = \lambda_\kappa/(N\mu)$}.
For the intermediate patient level
(\mbox{$\kappa = 2$}),
we have
\mbox{$\sigma_1 = r_{\text{hi}}$},
\mbox{$\sigma_2 = \rr$},
so that
\begin{equation}
\overline{L}\vphantom{L}^{\text{c}}_{2} = \frac{r_{\text{med}}}{(1-\rr)(1-r_{\text{hi}})} \,, \quad
\overline{W}\vphantom{W}^{\text{c}}_{2} = \frac{1}{N\mu}{\cdot}\frac{1}{(1-\rr)(1-r_{\text{hi}})} \,.
\label{LWmed}
\end{equation}
Observing that
\mbox{$\sigma_0 = 0$},
we obtain for the high-priority patients level
(\mbox{$\kappa = 1$}),
\begin{equation}
\overline{L}\vphantom{L}^{\text{c}}_{1} = \frac{r_{\text{hi}}}{1-r_{\text{hi}}} \,, \quad
\overline{W}\vphantom{W}^{\text{c}}_{1} = \frac{1}{N\mu}{\cdot}\frac{1}{1-r_{\text{hi}}} \,.
\label{LWhi}
\end{equation}
Results of this type were originally presented by \citet{NP:Cobham54} and
elaborated on by \citet{NP:Holley54}.
From \citep{NP:Kella85}, we also have the second moments
\begin{equation}
\overline{(W_\kappa^{\text{c}})^2} = \frac{2}{(N\mu)^2}{\cdot}
     \frac{1- \sigma_\kappa\sigma_{\kappa-1}}{(1-\sigma_\kappa)^2(1-\sigma_{\kappa-1})^3} \;,
     \quad \overline{(L_\kappa^{\text{c}})^2} = \overline{L}\vphantom{L}^{\text{c}}_\kappa
     + \lambda_\kappa^2 \overline{(W_\kappa^{\text{c}})^2} \;.
\label{SecondMom}
\end{equation}
The latter relationship can be seen to follow directly from (\ref{DistLittle}).
Specialization to high and intermediate priority patients
\mbox{$\kappa = 1,2$}
follows the same pattern as above for the means.

\section{Waiting-Time Distribution}
\label{Wait}
Let the RVs $L$ and $W$ represent the queue length and associated waiting time,
respectively, for a single priority level $\kappa$
of a non-preemptive Markovian priority queue.
According to the distributional form of Little's law \citep{NP:Bertsimas95,NP:Keilson88},
we can write
\begin{equation}
\langle e^{-r_\kappa W(1-z)}\rangle_W = \langle z^L\rangle_L \;,
\label{DistLittle}
\end{equation}
which gives the marginal queue-length PGF in terms of the waiting-time PDF as
\begin{equation}
G_L(z) = \int_0^\infty dt\, P_W(t) e^{-r_\kappa t(1-z)} \;.
\label{GtoP}
\end{equation}
It is assumed here that one is measuring time in units of $1/(N\mu)$,
which results from setting
\mbox{$N\mu= 1$},
and implies that
\mbox{$\lambda_\kappa = r_\kappa$}.
The relationship (\ref{GtoP}) may be inverted by means of an inverse Laplace transform,
and the Bromwich contour subsequently
deformed into one that wraps around the pole and cut singularities of the PGF.
Thus,
\begin{equation}
P_W(t) = r_\kappa\int_{\mathcal{C}_{\text{pol}} \cup \mathcal{C}_{\text{cut}}}
     \frac{dz}{2\pi i}\, e^{-r_\kappa t(z-1)} G_L(z) \;.
\end{equation}
For the wait-conditional problem, we can write
\begin{equation}
P^{\text{c}}_W(t) = \int_{\mathcal{C}_{\text{pol}} \cup \mathcal{C}_{\text{cut}}}
     \frac{dz}{2\pi i}\, f_\kappa(z;t)g^{(\kappa)}(z) \;,
\end{equation}
where $g^{(\kappa)}(z)$ is given by (\ref{g-kappa}) and
\mbox{$f_\kappa(z;t) \equiv r_\kappa e^{-r_\kappa t(z-1)}$}.

In the present application, we are interested in the intermediate patient priority level
(\mbox{$\kappa = 2$}).
In this case, the usual pole/cut decomposition
\mbox{$P^{\text{c}}_W(t) = P^{\text{c}}_{\text{pol}}(t) + P^{\text{c}}_{\text{cut}}(t)$}
reads \citep{NP:Zuk23C}
\begin{equation}
P^{\text{c}}_{\text{pol}}(t) = (1-\sigma)\left(1 - \frac{\sigma(1 - \sigma)}{r_{\text{med}}}\right)
     {\cdot}\frac{f_{\text{wt}}(1/\sigma;t)}{\sigma^2}
     {\cdot}\Theta(\sigma^2 - r_{\text{hi}}) \;,
\end{equation}
for the pole contribution, and
\begin{equation}
P^{\text{c}}_{\text{cut}}(t) = \frac{(1-\sigma)x_{\text{dif}}}{2\pi\sigma}
     \int_0^1 du\, \frac{\sqrt{u(1-u)}}{u+b}{\cdot}f_{\text{wt}}\left(x_{\text{dif}}(u+a);t\right) \;,
\end{equation}
for the cut contribution.
This result describes the waiting time for the aggregation of all intermediate-level patients.
For the waiting time of the intermediate-level ambulance-vehicle patients
in the presence of an APOT of size $M$,
\mbox{$P^{(M)}_{\text{med}}(t) \equiv P^{(M)}_{W}(t)$},
we must re-interpret the partial traffic intensity in (\ref {DistLittle}) as
\mbox{$r_2 \mapsto r_{\text{med}}^{\text{amb}} = pr_{\text{med}}$}
and consider
\begin{align}
\begin{aligned}
\chi P^{(M)}_{\text{med}}(t) &= \int_{\mathcal{C}_{\text{pol}} \cup \mathcal{C}_{\text{cut}}}
     \frac{dz}{2\pi i}\, f_{\text{wt}}(z;t)\frac{g(pz+q)}{z^M}
\\ &
     = p^{M-1}\int_{\mathcal{C}_{\text{pol}} \cup \mathcal{C}_{\text{cut}}}
     \frac{dz}{2\pi i}\, f_{\text{wt}}((z-q)/p;t)\frac{g(z)}{(z-q)^M} \;,
\end{aligned}
\end{align}
where
\mbox{$g(z) = (1-\sigma)/(\zeta_+(z) - \sigma)$}
and
\mbox{$f_{\text{wt}}(z;t) \equiv f_2(z;t)$}.
Here,
\mbox{$P^{(M)}_{\text{med}}(t) = P^{(M)}_{\text{pol}}(t) + P^{(M)}_{\text{cut}}(t)$},
with
\begin{equation}
\chi P^{(M)}_{\text{pol}}(t) = \frac{(1-\sigma)p^{M-1}\sigma^{M-2}}{(1 - \sigma q)^M}{\cdot}
     \left(1 - \frac{\sigma(1 - \sigma)}{r_{\text{med}}}\right)
     {\cdot}f_{\text{wt}}\left((1/\sigma - q)/p;t\right)
     {\cdot}\Theta(\sigma^2 - r_{\text{hi}}) \;,
\end{equation}
and
\begin{equation}
\chi P^{(M)}_{\text{cut}}(t) = \frac{(1-\sigma)(p/x_{\text{dif}})^{M-1}}{2\pi\sigma}
     \int_0^1 du\, \frac{\sqrt{u(1-u)}}{(u+a_q)^M(u+b)}{\cdot}
     f_{\text{wt}}\left((u+a_q)x_{\text{dif}}/p;t\right) \;.
\end{equation}
Explicit representations are given by
\begin{equation}
P^{(M)}_{\text{pol}}(t) = \frac{(1-\sigma)p^{M}\sigma^{M-2}}{\chi(1 - \sigma q)^M}{\cdot}
     \left(\sigma^2 - r_{\text{hi}}\right)
     \Theta(\sigma^2 - r_{\text{hi}}){\cdot}
     e^{-r_{\text{med}}t(1-\sigma)/\sigma} \;,
\end{equation}
and
\begin{equation}
P^{(M)}_{\text{cut}}(t) = \frac{\gamma(1-\sigma)(p/x_{\text{dif}})^{M}}{2\pi\sigma\chi}
     e^{-\gamma ct}\int_0^1 du\, \frac{\sqrt{u(1-u)}}{(u+a_q)^M(u+b)}{\cdot}
     e^{-u\gamma t} \;,
\end{equation}
where
\mbox{$\gamma \equiv r_{\text{med}}x_{\text{dif}} = 4\sqrt{r_{\text{hi}}}$}, \,
\mbox{$c \equiv \left(1 - \sqrt{r_{\text{hi}}}\right)^2/\gamma$}.
For the SF, we obtain
\begin{align}
\begin{aligned}
\bar{F}^{(M)}_{\text{pol}}(t) &= \frac{p^{M}\sigma^{M-1}}{\chi(1 - \sigma q)^M r_{\text{med}}}{\cdot}
     \left(\sigma^2 - r_{\text{hi}}\right)
     \Theta(\sigma^2 - r_{\text{hi}}){\cdot}
     e^{-r_{\text{med}}t(1-\sigma)/\sigma} \;, \\
\bar{F}^{(M)}_{\text{cut}}(t) &= \frac{(1-\sigma)(p/x_{\text{dif}})^{M}}{2\pi\sigma\chi}
     e^{-\gamma ct}\int_0^1 du\, \frac{\sqrt{u(1-u)}}{(u+a_q)^M(u+b)(u+c)}{\cdot} e^{-u\gamma t} \;.
\end{aligned}
\end{align}
The $L$-point quadrature rule reads
\begin{equation}
\bar{F}^{(M)}_{\text{cut}}(t) \simeq \frac{1}{\chi}\sum_{k=1}^L\frac{W_k e^{-(c + U_k)\gamma t}}
     {[(a_q + U_k)x_{\text{dif}}/p]^M(c + U_k)} \;,
\label{SFQuad}
\end{equation}
with the nodes and weights
\mbox{$U_k, W_k$}
as given by (\ref{UWGQ}).
Physical time units are recovered via the mapping
\mbox{$r_{\text{med}}t \mapsto \lambda_{\text{med}}t$}
in the scale invariants
\mbox{$P^{(M)}_W(t)dt$}
and
\mbox{$\bar{F}^{(M)}_W(t)$}
or, equivalently,
\mbox{$t \mapsto (N\mu)t$}.

\section{Exponential Ansatz}
\label{Ansatz}
Let us suppose that $P(n)$ is some queue-length PMF that is assumed to be geometrically
distributed, apart from a potential discontinuity at
\mbox{$n = 0$}.
Then, we can write
\mbox{$P(n) = At^{n-1}$}
for
\mbox{$n = 1,2,\ldots$}.
Since the $P(n)$ must sum to unity, we have
\mbox{$A = [1-P(0)](1-t)$}.
The mean queue length is given by
\begin{equation}
\bar{L} \equiv \sum_{n=0}^\infty nP(n) = \frac{A}{(1-t)^2} = \frac{1 - P(0)}{1-t} \;,
\end{equation}
from which it follows that
\mbox{$t = 1 - [1 - P(0)]/\bar{L}$}.
The exponential ansatz comprises applying this reasoning to the PMFs for the high-priority queue and the
aggregated intermediate-priority queue. The assumed form is exact for the high-priority patients
but only approximate for the intermediate-priority patients.

Now let us suppose that we partition the queue into a forward queue
comprising the first $M$ places, and an aft queue with the remainder.
Then, the queue-length PMF for the aft queue is given by
\begin{equation}
P_M(n) = \delta_{n0}\sum_{m=0}^M P(m) + (1 - \delta_{n0})P(n+M) \;,
\end{equation}
which has mean
\begin{equation}
\bar{L}_M \equiv \sum_{n=0}^\infty nP_M(n) = \sum_{n=1}^\infty P(n+M) = \bar{L}t^M \;.
\label{APOTShift}
\end{equation}
One may also observe that $P(0)$ can be determined with the help of the second moment
\mbox{$\overline{L^2} \equiv \sum_{n=0}^\infty n^2P(n)$},
according to
\begin{equation}
\overline{L^2} - \bar{L} = At\sum_{n=2}^\infty n(n-1)t^{n-2} = \frac{2At}{(1-t)^3} \;,
\end{equation}
which leads to
\begin{equation}
P(0) = 1 - \frac{2\bar{L}^2}{\bar{L} + \overline{L^2}} \;.
\label{PMFZero}
\end{equation}

Let $q$ denote the fraction of all intermediate-priority patients that are walk-ins.
Then the probability that ambulance-patient queue is empty is given by
\begin{equation}
P_{\text{med}}^{\text{amb}}(0) = \sum_{n=0}^\infty q^nP_{\text{med}}(n) \;.
\end{equation}
We combine this with the exponential ansatz
\begin{equation}
P_{\text{med}}(n) = [1 - P_{\text{med}}(0)](1-t)t^{n-1} \;, \quad
     t = 1 - [1 - P_{\text{med}}(0)]/\bar{L}_{\text{med}} \;,
\end{equation}
to find that
\begin{equation}
P_{\text{med}}^{\text{amb}}(0) = P_{\text{med}}(0) + [1 - P_{\text{med}}(0)]{\cdot}
     \frac{q(1-t)}{1 - qt} \;,
\end{equation}
where $P_{\text{med}}(0)$ is obtained by applying (\ref{PMFZero}) to the intermediate-priority level.
The result in (\ref{LAmbAnsatz}) of the main text is a direct consequence of the exponential ansatz,
followed by the shift (\ref{APOTShift}), where $M$ is the APOT size.
It only remains to make the identification
\mbox{$P_{\text{med}}^{\text{amb}} \equiv P_{\text{med}}^{\text{amb}}(0)$},
as given above.
\end{APPENDICES}

% For OPRE
% Endnotes command.
% \theendnotes

% Acknowledgments here.
% For OPRE.
% \ACKNOWLEDGMENT{The authors gratefully acknowledge useful discussions with Dr.~Stephen Bocquet.}
%
% For MOR.
\section*{Acknowledgments}
The authors gratefully acknowledge useful discussions with Dr.~Stephen Bocquet.

% References here (outcomment the appropriate case)

% CASE 1: BiBTeX used to constantly update the references
%   (while the paper is being written).
\bibliographystyle{informs2014} % outcomment this and next line in Case 1
\bibliography{APOTInforms} % if more than one, comma separated

% CASE 2: BiBTeX used to generate mypaper.bbl (to be further fine tuned)
%\input{mypaper.bbl} % outcomment this line in Case 2

%If you don't use BiBTex, you can manually itemize references as shown below.

%%%%%%%%%%%%%%%%%
\end{document}